\documentclass[aps,english,twocolumn,pra,longbibliography,notitlepage]{revtex4-1}
\usepackage{mathptmx}
\usepackage{graphicx}
\usepackage[T1]{fontenc}
\usepackage{enumerate}
\usepackage{amsthm}
\usepackage{amsmath}
\usepackage{amssymb}
\usepackage{blkarray}
\usepackage{xcolor}
\usepackage{tikz}
\usepackage{verbatim}
\usepackage{lipsum}
\usepackage{dsfont}
\usepackage[section]{placeins}
\usepackage{verbatim}
\usepackage{lipsum}
\usepackage{dsfont}
\newcommand{\newc}{\newcommand}
\newc{\beq}{\begin{equation}}
\newc{\eeq}{\end{equation}}
\newc{\kt}{\rangle}
\newc{\br}{\langle}
\newc{\beqa}{\begin{eqnarray}}
\newc{\eeqa}{\end{eqnarray}}
\newc{\Tr}{\mbox{Tr}}
\newc{\ep}{\text{ep}}
\newc{\ovl}{\overline}
\newc{\longra}{\longrightarrow}
\newc{\ot}{\otimes}
\newc{\evn}{\mathcal{E}_{vn}}

\newc{\ca}{{\cal A}}
\newc{\cb}{{\cal B}}
\newc{\ta}{{\tilde A}}
\newc{\tb}{{\tilde B}}
\newc{\colr}{{\color{blue}}}
\newc{\nn}{\nonumber \\}

\newtheoremstyle{mystyle}
  {}
  {}
  {\normalfont}
  {}
  {\itshape}
  {.}
  { }
  {}
\theoremstyle{mystyle}

\makeatletter
\let\Hy@backout\@gobble
\makeatother

\begin{document}

\title{Probing the randomness of ergodic states: extreme-value statistics in the ergodic and many-body-localized phases }

\author{Rajarshi Pal }
\affiliation{Department of Physics, Sungkyunkwan University, Suwon 16419, Korea.}

\author{Arul Lakshminarayan}
\affiliation{Department of Physics, Indian Institute of Technology Madras, Chennai, India 600036.}

\begin{abstract}
The extreme-value statistics of the entanglement spectrum in disordered spin chains possessing a many-body localization transition is examined. It is expected that eigenstates in the metallic or ergodic phase, 
behave as random states and hence the eigenvalues of the reduced density matrix, commonly referred to as the 
entanglement spectrum, are expected to follow the eigenvalue statistics of a trace normalized Wishart ensemble. In particular, the density of eigenvalues is supposed to follow the universal Marchenko-Pastur distribution. We find deviations in the tails  both for the disordered XXZ with total
$S_z$ conserved in the half-filled sector as well as in a model that breaks this conservation. A sensitive measure of deviations is provided by the largest eigenvalue, which in the case of the Wishart ensemble after 
appropriate shift and scaling follows the universal Tracy-Widom distribution. We show that for the models considered, 
in the metallic phase, the largest eigenvalue of the reduced density matrix of eigenvector, instead follows the generalized extreme-value
statistics bordering on the  Fisher-Tipett-Gumbel distribution indicating that the correlations between eigenvalues are much weaker compared
to the Wishart ensemble. We show by means of distributions conditional on the high entropy and normalized participation ratio of eigenstates that the conditional entanglement spectrum still follows generalized extreme value distribution. In the deeply localized phase we find heavy tailed distributions and L\'evy stable laws in an appropriately scaled function of the largest and second largest eigenvalues. The scaling is motivated by a recently developed perturbation theory of weakly coupled chaotic systems.  
\end{abstract}

\maketitle 

\section{Introduction}

The phenomenon of Anderson localisation \cite{And58} has been found to survive interactions \cite{Basko06,Imbrie16} and a flurry of research activity has been devoted to understanding and characterising this transition to a localized phase from a delocalized or ergodic one. This phenomenon widely referred to as many-body localisation (MBL) is fundamentally interesting as such systems generically break ergodicity and fail
to thermalise--thus lying beyond the scope of statistical mechanics. Also, MBL occurs throughout, and especially the middle of the spectrum implying that it is an infinite temperature quantum phase transition, different from usual quantum phase transitions studied at zero temparature\cite{HN15,Huse14,BPM12,PH10}. These facts combined have significant practical implications for quantum transport \cite{Basko06} and information storage \cite{YLV15,Demlu14}. Experimental advances have allowed the controlled observation of MBL
phenomena \cite{Schreiber15}, further driving interest. 
         
           Ideas from quantum information have played an important role in the development of the understanding of the ergodic-localized transition. Quantum entanglement, a topic of much importance in quantum information theory, has also gained
relevance in quantum many-body physics in the past few years \cite{Amico08,Eisert10}. In particular, the entanglement entropy provides a wealth of information about physical states, including novel ways to classify states of matter that do
not have a local order parameter \cite{Kitaev06,Wen03}. Many-body eigenstates of thermalizing systems exhibit an entanglement entropy that scales with the volume of the subregion being considered, while many-body eigenstates
of many-body localized systems display a boundary law
scaling (with possible logarithmic corrections). The study of entanglement entropy and its dynamics has played a crucial
role in the elucidation of the properties of the localized and thermal or ergodic phases. Indeed, studies of entanglement entropy \cite{Pollman12,Pollman14} provided the first clues as to the emergent integrability of
the localized phase. The differences in the nature of multipartite entanglement in the thermal and localized phase is also complementarily reflected in quantities such as a concurrence that measure two qubit entanglements \cite{Bera2016}.  However, entanglement entropy captures only a small part of the full
entanglement structure of a system. Much greater information is contained in the entanglement spectrum \cite{Haldane08}, from which
the entanglement entropy, and much more, maybe extracted.   

               In the context of MBL, the entanglement spectrum has been studied in some recent papers
               \cite{Chamon15,Nand16,Serbyn16,Parisi17} and power laws have been found in disorder averaged Schmidt eigenvalues plotted 
               against the eigenvalue order.       
In the ergodic phase, different statistical properties of energy levels, such as the ratio of nearest neighbor spacings (\cite{Atas2013})  have been shown to correspond to that of one of the canonical random matrix ensembles, the GOE (Gaussian orthogonal ensemble) \cite{PH10}. In fact going beyond short range correlations the number variance \cite{Mehta, Leshousches89} between levels was computed in \cite{Bertrand16} and was shown to be consistent with the corresponding random matrix theory (RMT) formulae till about $100$ mean-level spacings. This together with the claim that the limiting density of eigenvalues of the reduced density matrices of the eigenstates is identical to the Marchenko-Pastur \cite{Chamon15} would indicate that deep in the ergodic, delocalized phase, the system is well described by standard random matrix ensembles.
A consequence of this is that the eigenstates should be random states and the reduced density matrix (of say $k$ spins out of a total of $L$) belong to the so-called trace-constrained Wishart ensemble $MM^{\dagger}/\Tr(MM^{\dagger})$, where $M$ is a random matrix of dimension $2^k \times 2^{L-k}$, whose entries are zero-centered independent normal random numbers. The average entanglement entropy is the von Neumann entropy of such an ensemble and is given by the Page value \cite{Page93}.

It is known that the largest eigenvalues of the Wishart ensemble after a suitable shift and scaling satisfies the universal Tracy-Widom 
distribution \cite{Tracy96}, a deviation from the classical extreme value statistics due to the strong correlations of the eigenvalues. On the other hand it is also well known that if a set of random variables are independent and identically distributed then for appropriately rescaled variables
there are three possible limiting universal distributions for the extreme maximal events: the Fr\'echet, Fisher-Tipett-Gumbel, and Weibull distributions \cite{FisherTippett1928,Gumbel04}. Respectively, they arise depending on whether the tail of the density is a power law, or
faster than any power law, and unbounded or bounded. If there are correlations, then it is known that these universal distributions are still valid and reached for sufficiently fast decay of autocorrelations \cite{Leadbetter88}. Thus that the Tracy-Widom law differs from these distributions is the consequence of the peculiar strong correlations present in the eigenvalues of random matrices. 

The results of the present paper indicate that the distribution of the maximum of the entanglement spectrum deviates significantly from the 
Tracy-Widom distribution and instead fits quite well a Fisher-Tipett-Gumbel distribution. This is seen both in a disordered XXZ model with total $S_z$ conserved, which has become a 
standard model for studying the transition and also in a model with an extra field breaking it which shows the two-component structure \cite{Chamon15}. This indicates that correlations in the entanglement spectrum are not Wishart like even deep in the delocalized phase, and perhaps shows signs of pre-localization \cite{Luca13}. Most eigenstates in the metallic state have high entanglement and are delocalized as reflected by the participation ratio. We look at conditional distributions of the maximum eigenvalue of the density matrices filtering only high entropy and participation ratio states and they also follow generalized extreme value distribution (GEV) distributions and not the Tracy-Widom statistic. This is seen to hold even as we start moving towards the transition point but away from it towards the MBL phase, interestingly, the distribution becomes more Fr\'echet-like. 
Even when the maximum distribution starts deviating significantly from GEV, we show that the conditional distribution still follows GEV, 
signaling significant heterogeneity in the spectrum. In the localized phase, we find power laws in the scaled distribution of the largest and 
second largest eigenvalues. These are  quite distinct from
the power laws in the disorder averaged entanglement spectra itself plotted against the eigenvalue order \cite{Serbyn16}.
  
Previous studies on extreme eigenvalues of the reduced density matrices were mostly in the context of strongly chaotic bipartite systems. Remarkable exact results, beyond the aymptotic universal distributions, have been obtained within random matrix theory (RMT) for both the maximum and the minimum eigenvalues and these have been compared with conductance fluctuations in chaotic cavities and entanglement spectra of systems such as coupled kicked tops \cite{Majumdar2008,MajumdarBook,KAT2008,Vivo2011,Kumar2017,    Forrester2019}. A study of the transition in the distribution of the largest eigenvalue of the reduced density matrix of eigenstates ofchaotic bipartite systems as a function of interaction, showed that the largest eigenvalue showed deviations from statisticality 
even when the interactions are strong enough for other statistics to show random matrix behavior \cite{Tomsovic2018}.

\section{Preliminaries: extreme-value statistics and entanglement} 

This section collects well-known details about extreme-value
statistics that are of relevance to this paper, including a discussion of 
entanglement in random states and the implications for the extreme
eigenvalues of the reduced density matrices.

\subsection{Extreme-values of independent or weakly correlated random variables}
Let $X_1,X_2,.....X_N$ be $N$ observations of an identical independent random variable $x$ with density $p(x)$. Under very general conditions we know from the central limit theorem that the mean $ \sum_iX_i/N$ is normally distributed
for large $N$. However, the extremes  $(X_{\max})_N=\max \{X_1,X_2,...X_N\}$  and $(X_{\min})_N$ defined similarly are {\emph{not}} normally distributed, but still show universal behavior. 
Let,
\begin{multline*}
\text{Prob}((X_{\max})_N <x)=F_N(x)= \\ \text{Prob}(X_1<x,X_2<x,...X_N<x)
= \left(\int_0^x p(x)dx \right)^N.  
\end{multline*}
For large $N$, after a shift and change of scale $y=(x-a_N)/b_N$ , $F_N(y)$ tends to one of three universal distributions, depending on the tail of the 
density $p(x)$ \cite{FisherTippett1928,Gumbel04}. These are 
\begin{enumerate}[(i)]
\item {\emph{Fisher-Tipett-Gumbel:}} $\exp{(-\exp{(-y)})}$, if the tail of $p(x)$  decays as $\sim e^{-x^{\delta}}$, $\delta>0$, faster than a power law,
\item {\emph{Weibull:}} $\exp{(-(-y)^{\alpha})}$ for $y \leq 0$ and $0$ otherwise, if $x$ is bounded above, and the tail of $p(x)$ decays as 
$\sim |x|^{-1-\alpha}$, $\alpha>0$, 
\item {\emph{Fr\'echet:}} $\exp{(-y^{\alpha})}$ for $y \geq 0$ and $0$ otherwise, if $x$ is bounded below, and the tail of $p(x)$ decays as 
$\sim x^{-1-\alpha}$, $\alpha>0$. 
\end{enumerate}
The shift, which is the typical size of the extreme value, and scale paratameters $a_N$ and $b_N$ depend on the sample size $N$ and $\alpha$ or $\delta$, in particular for the 
Fisher-Tipett-Gumbel law, $a_N$, which follows on requiring that 
$\int_{a_N}^{\infty} p(x) dx=1/N$, is $\sim (\log N)^{1/\delta}$. 
These distributions can be expressed in a combined way through the generalized extreme value distribution (GEV) with the cumulative distribution function,
\begin{equation}
\label{eq:gev}
\begin{split}
F(y;\xi) &=  \exp{(-(1- \xi y)^{1/\xi})} \; \mbox{for} \; \xi \ne 0  \\  
&=  \exp{(-\exp{(-y)})} \; \mbox{for} \; \xi=0,  
\end{split}
\end{equation}
with $\xi$ being the shape parameter, $\xi=0$,  $\xi<0$ and $\xi>0$ corresponding respectively to the Gumbel, Fr\'echet and Weibull families. 

If the random variables are correlated, then weak correlations in the
sense that $\br x_i x_j \kt -\br x_i \kt \br x_j\kt$ is say exponentially small $\sim e^{-|i-j|/\xi}$ with $\xi \ll N$ then the 
extremes still follow one of the three classical extreme distributions.
The maximum intensity in random states, for example, is an exactly solvable case of weakly correlated random variables that limit to the 
Fisher-Tipett-Gumbel distribution \cite{AL08}. One well-known case where
strong correlations lead to a limiting distribution, the Tracy-Widom distribution, different from
the above three classical ones, are the eigenvalues of random matrices
\cite{Tracy96}. However, for our purposes, we are interested in the
singular values of random matrices which are closely connected to entanglement and the largest values also follow the Tracy-Widom distribution \cite{Johnstone01}.

\subsection{ Extreme-value statistics, entanglement and the Wishart ensemble}

Let us begin by introducing some of the properties of 
the Wishart ensemble of random matrix theory which is the appropriate ensemble for modelling statistics of the entanglement spectrum.  
We will discuss the real Wishart ensemble throughout, as the Hamiltonians we consider have time-reversal anti-unitary symmetry \cite{Nand16}. 

Let, $|\Phi \rangle$ be a state belonging to the tensor product space ${\mathcal{H}}_1 \otimes {\mathcal{H}}_2$ with $\text{dim}({\mathcal{H}}_1)= n_1$,
$\text{dim}({\mathcal{H}}_2)= n_2$  and $n_1 \leq n_2$. In our case, since we will be considering entanglement across two $L/2$ partitions ofspin-1/2 chains, we will have
$n_1=n_2=2^{L/2}$. 
Let, $\{|i \rangle, | j \rangle \} $ be an orthonormal basis in ${\mathcal{H}}_1$ and ${\mathcal{H}}_2$  respectively. We have with respect to this
basis the state  $|\Phi \rangle$  and its Schmidt decomposition,
\begin{equation}
|\Phi  \rangle = \sum_{i=1}^{n_1}  \sum_{j=1}^{n_2} c_{ij}|i \rangle \otimes | j \rangle= \sum_{i=1}^{n_1} \sqrt{\lambda_i} |i' \rangle_1 |i' \rangle_2.  
\end{equation}
The reduced density matrix of the subsystems are given by $\rho_1=\Tr_2(|\Phi \rangle \langle \Phi|)$ and $\rho_2=\Tr_1(|\Phi \rangle \langle \Phi|)$.  It follows that
 Schmidt coefficients $\lambda_i$ are eigenvalues of $\rho_1=CC^{\dagger}$, or equivalently $\rho_2=(C^{\dagger}C)^T$\cite{Nielsen11}, 
 where $C$ is the ``coefficent matrix" with elements $c_{ij}$. 

The state is unentangled if and only if $\lambda_1=1$ (assuming the Schmidt coefficients are ordered and hence all other eigenvalues are $0$), and the Schmidt decomposition
gives the states of the individual subsystems. Otherwise $\lambda_2> 0 $ and the Schmidt decomposition consists
of at least two terms. For maximally entangled states, $\lambda_j = 1/n_1$  for all $j$. The entanglement entropy
in the state $|\Phi  \rangle$ is the von Neumann entropy of the reduced density matrices, $S= -\Tr(\rho_1 \log{\rho_1})= -\Tr(\rho_2 \log{\rho_2})$.   
 If $S = 0$, then the state is unentangled, while a maximally entangled state has $S = \log{n_1}$ . 

Hamiltonians are modeled as random matrices from the GOE in the so called ergodic phase of many-body systems. If the coefficients $c_{ij}$ come from an eigenvector of a typical GOE matrix the induced probability distribution on the Schmidt eigenvalues $\lambda_i$ and the consequences for entanglement are well-known. \cite{SommersZyczkowski2001, BZBook}
The  eigenvectors of a matrix from a GOE of dimension $n$, are only constrained by 
normalization and the joint distribution of their components is hence given by,
\begin{equation}
\label{eq:jpdfeigv}
P(x_1,x_2,...x_n)= \frac{\Gamma(n/2)}{\pi^{n/2}} \delta \left(\sum_i x_i^2-1\right).
\end{equation}
Hence, the distribution of $c_{ij}$ is same as that of $[M_{ij}]/\sqrt{\Tr(MM^{\dagger})}$ with $M$ being an unstructured
matrix with all elements {\it i.i.d.} zero mean normally distributed numbers, the standard orthogonal ``Ginibre ensemble'' \cite{Ginibre65}. Thus the reduced density matrices are given by the ensemble of random matrices,
\begin{equation}
\label{eq:ginibre}
\rho = \frac{MM^{\dagger}}{\Tr(MM^{\dagger})}.  
\end{equation}
These are the so called trace constrained Wishart ensemble of random matrix theory.
 
The joint probability density function (j.p.d.f.) of $\lambda_i$, the eigenvalues of $\rho$, is
\begin{equation}
\begin{split}
&P(\lambda_1,\cdots,\lambda_{n_1}) = B_{n_1,n_2} \delta\left(\sum_{i=1}^{n_1} \lambda_i -1 \right)\\ 
&\prod_{i=1}^{n_1} \lambda_i^{\frac{\beta}{2}(n_2-n_1+1)-1} \prod_{j<k} |\lambda_j - \lambda_k|^{\beta},
\end{split}
\end{equation} 
where $\beta =1,2$ corresponds to real or complex entries of $M$ and $B_{n_1,n_2}$ is a normalization constant known explicitly \cite{Mehta}. 
The symmetry or Dyson index is $\beta=1$ for time-reversal symmetric systems such as considered in this paper. There are two sources of correlations in the $\lambda_i$, the delta function is a much weaker source of correlation and is identical to that of eigenfunction components as both originate from normalization \cite{AL08}. The strong and peculiarly RMT correlations arise from the Vandermonde determinant factor involving the product of the differences of every pair of eigenvalues. In the absence of the j.p.d.f. for the eigenvalues of the reduced density matrices of the eigenstates of many-body systems, we want to investigate if the consequences of the strong correlations for the largest eigenvalue are present in the physical systems.
 
However, before delving into this question, we also investigate the 
average entanglement and the density of states (of $\lambda_i$), which follow from the j.p.d.f.. While an exact result is known for $\beta=2$:
(the Page formula) \cite{Page93},
\begin{equation}
\langle S \rangle = \sum_{k=n_2 + 1}^{n_1 n_2}\frac{1}{k} - \frac{n_1-1}{2n_2} \approx \ln n_1 - \frac{n_1}{2n_2}, 
\end{equation}
the asymptotic result for large $n_1$ and $n_2$ is valid for both $\beta=1$ and $\beta=2$. 

\subsubsection{Marchenko-Pastur Law} 
The average density of the Schmidt eigenvalues is obtained by integrating out all variables except one. For $n_2 \geq n_1 \gg 1 $, with fixed ratio $Q=n_2/n_1 \geq 1$, the limit of the density of scaled eigenvalues $\tilde{\lambda_i}=\lambda_i n_1$ is given by the Marchenko-Pastur law,

\begin{equation*}
\rho^Q_{MP}(x) = \frac{Q}{2 \pi} \frac{\sqrt{(x_+ - x)(x - x_{-})}}{x},  x_{-} \leq x \leq x_+,
\end{equation*}
and $0$ otherwise. The distribution is in the finite support $[x_{-}, x_+]$, where $x_{\pm} = 1 + 1/Q \pm 2/\sqrt{Q}$. 

For the case $Q=1$, especially relevant for the numerical results presented, the distribution is given by,
\begin{equation}
\label{eq:MP} 
\rho_{MP}(x) = \frac{1}{2\pi} \sqrt{\frac{4-x}{x}}, \; 0 \leq x \leq 4, 
\end{equation}
and zero otherwise. The distribution thus diverges at the origin. The Marchenko-Pastur law is in fact a 
universal distribution for ensembles of correlation matrices, irrespective of the exact distribution of matrix
elements as long as it has a finite moments of sufficiently larger order \cite{TaoVu2012}. 
The moments of the Marchenko-Pastur distribution, $M_n$ are given by the Catalan  numbers, \[ \br x^n \kt =C_n=\frac{1}{n+1} {2n \choose n}, \] thus 
$M_n^{1/n}  \rightarrow 4$ as $n \rightarrow \infty$ \cite{Haake2001}.

\subsubsection{Distribution of maximum eigenvalue}

As mentioned earlier, the distribution of the maximum eigenvalue of a Wishart ensemble after suitable centering and scaling follows the Tracy-Widom distribution \cite{Tracy96} which has no simple closed-form. For the orthogonal case, it can be defined implicitly by the Hastings-McLeod 
solution to the second Painlev\'e equation, \cite{Tracy96}. 
For our purposes, since the reduced density matrix corresponding to a random state has unit trace, we need to adapt the results of the Wishart ensemble to the trace constrained one. This was done in \cite{Nechita07}, for complex
Wishart matrices. Adapting the methods used there with the results of real Wishart matrices in \cite{Johnstone01} we obtain the same center and scaling in the large $n$ limit as complex matrices, namely a centering or shift of  $4/n$ and a scaling 
equal to  $2^{\frac{4}{3}}n^{-\frac{5}{3}}$. This is obtained as follows. 

We have our reduced density matrix $\rho=W/S$ (with $W=M^{\dagger}M$ and $S=\Tr (W)$, Eq.~(\ref{eq:ginibre})) and thus
$\lambda_{\max}(\rho)=\lambda_{\max}(W)/S$. 
Now, from \cite{Johnstone01} we know that mean of the distribution of  $ \lambda_{\max}(W)$,  $\langle \lambda_{\max}(W) \rangle$ goes as 
$(\sqrt{n-1} + \sqrt{n})^2$ while  mean of the distribution of trace of $W$ goes as $\langle S \rangle \sim n^2$. Hence, 
we approximately have the mean of $ \lambda_{\max}(\rho)$ equal to,  
$\langle  \lambda_{\max}(\rho) \rangle  \sim \frac{(\sqrt{n-1} + \sqrt{n})^2}{n^2}= \frac{4}{n},$ for large $n$. 
Again from \cite{Johnstone01} we know that the standard deviation of the distribution of $\lambda_{\max}(W)$
, \[\sigma_{{\lambda_{\max}(W)}} \sim (\sqrt{n-1} + \sqrt{n})(\frac{1}{\sqrt{n-1}}+ \frac{1}{\sqrt{n}})^{\frac{1}{3}}. \]  
As, the fluctuation in the largest eigenvalue is much greater than the fluctuation in the sum of eigenvalues, i.e, 
\[\frac{\sigma_{\lambda_{\max}(W)}}{\langle \lambda_{\max}(W) \rangle } \gg \frac{\sigma_{S}}{\langle S \rangle} \] we approximately have, 
$\sigma_{\lambda_{\max}(\rho)}= \frac{\sigma_{{\lambda_{\max}(W)}}}{\langle S \rangle}= 2^{\frac{4}{3}}n^{-\frac{5}{3}}$ for large $n$.

\section{Numerical results of the statistics in two spin models}

{\emph{Model:}} We consider the following model for MBL, which is an XXZ spin-$\frac{1}{2}$ chain of $L$ spins with a random $z$ field and a small constant $x$ field. 
\begin{equation}
H= \frac{1}{2}\sum_{i=0}^{L-1} \left( \sigma_i^x \sigma_{i+1}^x +  \sigma_i^y \sigma_{i+1}^y + \Delta \sigma_i^z \sigma_{i+1}^z \right) + \sum_i h_i \sigma_i^z + \Gamma \sum_i \sigma_i^x
\end{equation}
$h_i$ chosen to be \emph{i.i.d.}  uniformly random in $[-W,W]$. We consider both $\Gamma=0$ and $\Gamma=0.1$. For the $\Gamma=0$ case, total $S_z$ is conserved and we restrict ourselves to the half-filled sector with open boundary conditions.  
Since, we are interested in studying how {\emph{random}} the model truly is in the ergodic phase, we need to break the   $S_z$  conservation without affecting the transition too much to see the effect of this conservation on the randomness. 

We will refer to the model with $\Gamma=0$ and $\Gamma=0.1$ respectively as $H_1$ and $H_2$. The model $H_1$ has been extensively studied and is believed to capture all essential properties of the MBL
phase and the localization transition. For example, it is known that the model supports a MBL phase at strong disorder, an ergodic phase at weaker disorder, and an 
integrable point at zero disorder.  For $\Delta=1$ the transition was estimated to happen around $W \approx 3.5$ based on a variety of probes like the level statistics \cite{PH10,Luitz15,Serbyn16} and fluctuations of entanglement entropy \cite{Pollman14}. 
Models with random fields along different directions have also been studied for example in \cite{Nand16,Nandkishore16} and MBL transition  with similar properties as $H_1$ was found.   

Throughout the paper we have considered the middle one-third eigenstates for data. 
We have checked that there is no significant difference in the distributions we have computed if we instead choose a single eigenstate
from the middle of the spectrum and many more disorder realizations. $500$ disorder realizations  are chosen for the $L=14$, $H_1$ model and $100$ disorder realizations for $L=12$, $H_2$ model.  

  Let $X$ be a discrete random variables with outcomes $1,2,...n$ and let $p_i=P(X=i)$. The R\'enyi entropy of order  $\alpha$, where $\alpha \geq 0$ and $\alpha \ne 1$ is
  defined as \[H_\alpha(X)= \frac{1}{1-\alpha} \log{\left(\sum_{i=1}^n p_i^\alpha \right)}. \] In the limit of $\alpha \rightarrow \infty$ the Renyi entropy converges to the min-entropy,
\begin{equation}
H_{\infty}(X)= \min_i \{-\log{p_i}\} = -\log {\max_i p_i}.  
\end{equation} 
The min-entropy, as its name indicates, is the smallest of the R\'enyi entropies   and is the most conservative estimate of the information content of a random variable. In this paper we study the maximum of the entanglement spectrum, which essentially provides us  
with the min-entropy of entanglement.

\subsection{Deviations from the Marchenko-Pastur distribution}
\begin{figure}
\includegraphics[width=0.5\textwidth]{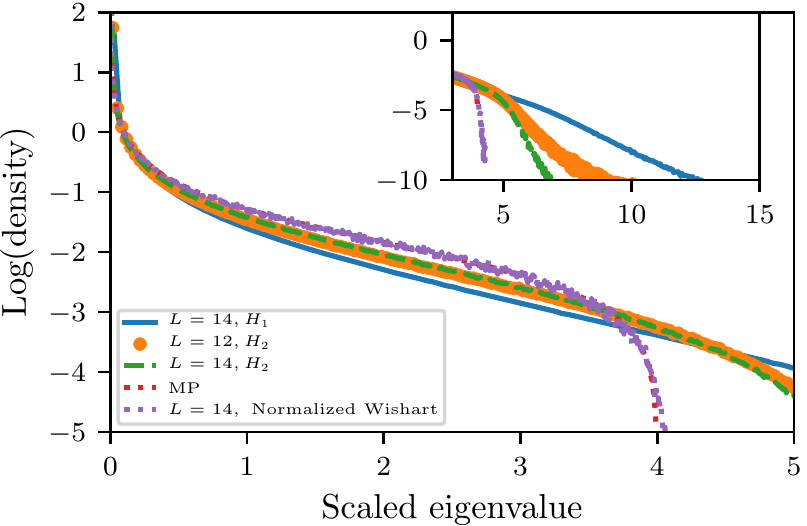}
\caption{Log of the scaled eigenvalue distribution for different $L$ for $H_2$ and $H_1$ for $L=14$, $W=0.5$, MP refers to the Marchenko-Pastur distribution in Eqn.~ (\ref{eq:MP}).The tail is shown in the inset.}
\label{fig:2}
\end{figure}

We first compare the average density of the entanglement spectrum  with the Marchenko-Pastur (MP) distribution. This is computed 
for $W=0.5$ and shown for different $L$ for $H_2$ and $H_1$  in Fig.~(\ref{fig:2}).
In the figures, scaled eigenvalue refers to the eigenvalues multiplied by the dimension $2^{L/2}$. 
For $L=14$, in the $H_2$ model $20$ disorder realizations have been used. 

As is clear, breaking the total $S_z$
conservation brings the distributions considerably closer to the MP distribution. While the density approaches the MP distribution for larger values of $L$, the tails show that the limiting distribution is perhaps close to  MP,  but different. Similar observations were reported in \cite{Parisi17}. The Marchenko-Pastur has finite support from 0 to 4, but the distributions obtained from the data show 
exponential tails. Of course, the hard bound is obtained in the asymptotic $L \rightarrow \infty$ limit, however the softening of this for finite $L$ \cite{KAT2008} is too small in comparison to the tails observed here.   


\begin{figure}
\includegraphics[width=0.5\textwidth]{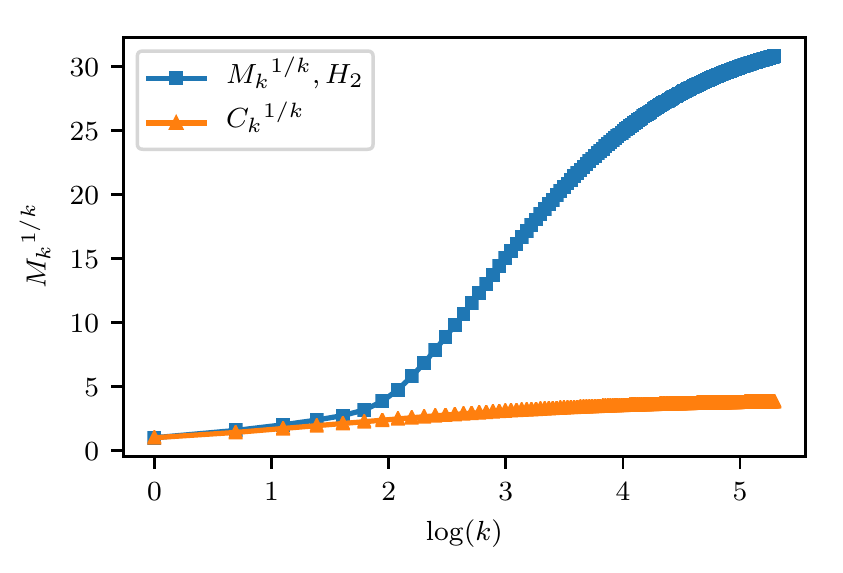} 
\caption{Plot of ${M_k}^{1/k}$ with $\log(k)$ for $L=12, H_2$ and  $W=0.5$ (squares). The other curve (triangles) depicts ${C_k}^{1/k}$ with $\log(k)$ and 
saturates to $4$.}
\label{fig:3}
\end{figure}

This is also reflected, as it should be, in the deviations of the moments from that of the MP distribution. The deviations from this are shown in 
Fig. (\ref{fig:3}) for $H_2$, with the blue and orange curves representing respectively ${M_k}^{1/k}$ of the distribution obtained for $L=12, W=0.5$ and that of the Marchenko-Pastur distribution.  

\subsubsection{Deviation from gaussianity and normalized participation ratio}
\begin{figure}
\includegraphics[width=0.5\textwidth]{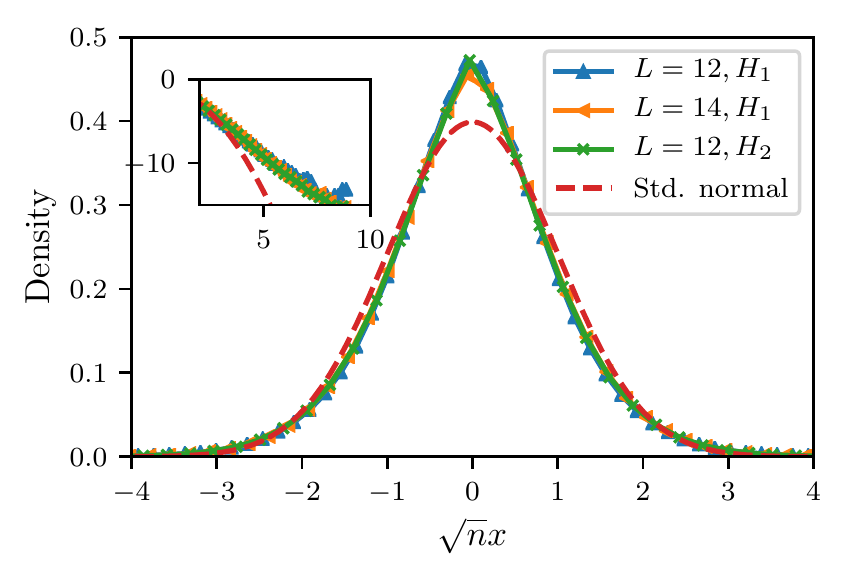} 
\caption{Distribution of scaled eigenvector components $\sqrt{n}x$  for  $L=14, H_1$ and $L=12,H_2$ and disorder strength $W=0.5$. The inset shows the tail, 
the y-axis of the inset represents the logarithm of the density.}
\label{fig:7}
\end{figure}

\begin{figure}
\includegraphics[width=0.5\textwidth]{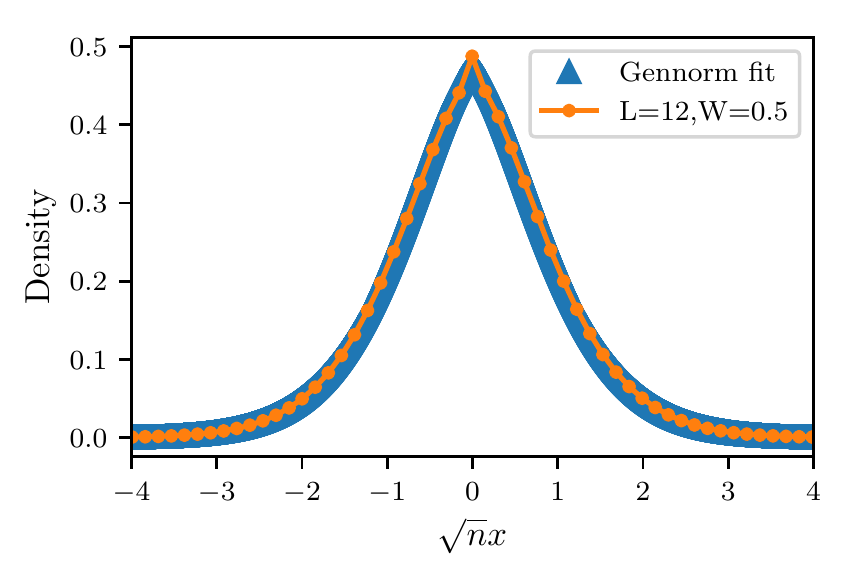} 
\caption{Distribution of scaled eigenvector components $\sqrt{n}x$ for $L=12,H_2$ and $W=0.5$, compared to a generalized normal distribution as in Eq.~(\ref{Eq:vectordistr}).}
\label{fig:7p5} 
\end{figure}

While looking for  deviations shown by the entanglement spectrum statistics from the predictions of Wishart ensemble, it is natural to ask
if the normalized components of eigenvectors ($\sqrt{n}x_i$, see Eqn. (\ref{eq:jpdfeigv})) are themselves normally distributed, to begin with. While this would be the case if the GOE ensemble were to apply,
although this is not a necessary condition for the MP distribution, due to its universality.
As seen  in 
Fig.~(\ref{fig:7}), while the distributions approach Gaussian for increasing $L$, the fact that the logarithm of the distribution has near linear
rather than quadratic tails, as seen in the inset, indicates that the limiting distribution is perhaps different. 
Interestingly, as seen in Fig.~(\ref{fig:7p5}) the distribution of the normalized coefficients of the eigenvectors fits an exponential power
distribution with p.d.f. 
\beq 
\label{Eq:vectordistr}
P(x)= \frac{\beta}{2 \alpha \Gamma(1/ \beta) } \exp{\left(-\left|\frac{x}{\alpha}\right|^{\beta}\right)}
\eeq
 quite well. This is a generalization of the normal distribution ($\beta=2$) with an additional shape parameter 
$\beta$. The $\beta$ for the fits in the figure  is to the leading order $1.5$ and the scale factor $\alpha=1.2$.

The deviation away from Gaussianity can be measured by the kurtosis of the distribution of scaled eigenvector coefficients. As the mean of the distribution is still close to  zero, when computed for a single 
eigenvector, the kurtosis is equal to $\frac{1}{n} \sum_i (\sqrt{n} x_i)^4$, which is just the normalized inverse participation ratio , $n\sum_ix_i^4$ with respect to the product basis in which $H_1$ and $H_2$ are diagonalized. Here we compute the inverse of this quantity, the normalized participation ratio (PR) which is the inverse of the kurtosis for different states. This mean value of the PR is known to be equal to $1/3$ for the GOE ensemble \cite{Pandey81}.

\begin{figure}
\includegraphics[width=0.4\textwidth]{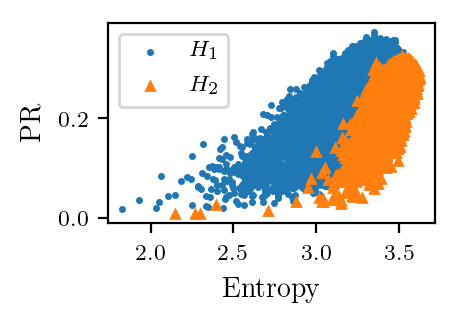} 
\caption{ Entanglement entropy \emph{vs} participation ratio for  $L=12, H_1$,  500 realizations and $L=12,H_2$,100 realizations and $W=0.5$.}
\label{fig:8}
\end{figure}

Also, random states are highly entangled, with the average entropy given by the Page value $ \approx \ln{ n} - \frac{1}{2}$ ~\cite{Page93}. 
In Fig.~(\ref{fig:8}) we show the entropy vs. PR plots for $H_1$ and $H_2$ for $L=12$.  
The strong correlation between entropy and PR is clear. Thus most eigenstates are highly entangled with a participation ratio close to GOE.
The mean PR for $H_1$ and $H_2$ are respectively, $0.259$ and $0.264$. The mean entropies are respectively
$3.27$ and $3.5$, while the Page value $\approx 3.66$. Also as is clear from  Fig.~(\ref{fig:8}), in the model with total $S_z$ conservation 
the PR and entropy have  lower variance and are closer to random values as expected.   

\subsection{Deviations of the maximum from the Tracy-Widom distribution}
 Following the adaptations mentioned before, here we compare the maximum eigenvalue data after using a center and scaling respectively 
 of   $\frac{4}{n}$ and $2^{\frac{4}{3}}n^{-\frac{5}{3}}$ with $n=2^{L/2}$. We have,
\begin{equation}
\label{eq:scaledlambmax}
\lambda_1' = \frac{(\lambda_1 - \frac{4}{n})}{ 2^{\frac{4}{3}}n^{-\frac{5}{3}}} . 
\end{equation}
 
\begin{figure}
\includegraphics[width=0.5\textwidth]{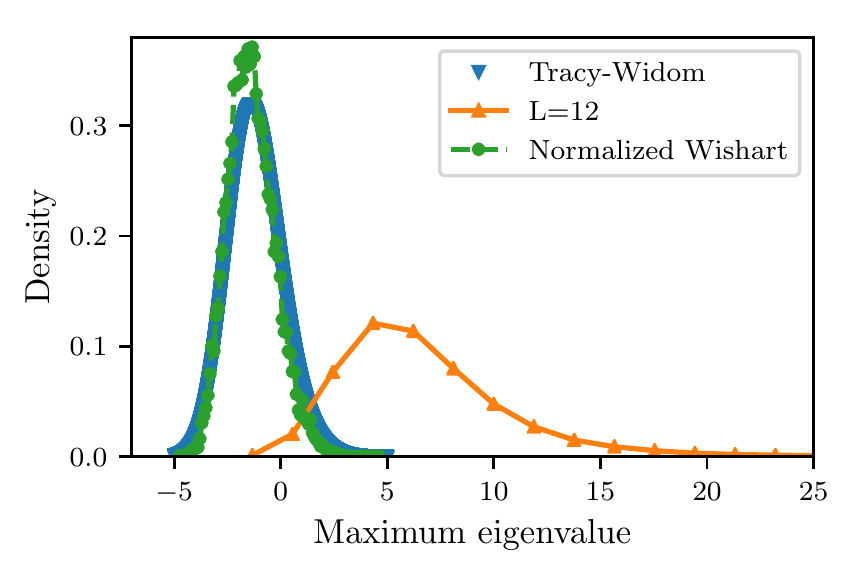}
\caption{Comparison of the distribution of the shifted-and-scaled maximum eigenvalue, $\lambda_1'$, Eq.~(\ref{eq:scaledlambmax}), with the Tracy-Widom distribution for $L=12,H_2$ and $W=0.5$. Shown for comparison is also the case of a (trace normalized) Wishart
ensemble of dimension $2^{L/2}=64$.}
\label{fig:5}
\end{figure}

In order to compare for finite size effects, we also plot data from a trace normalized Wishart ensemble of the same dimensions. 
This is shown in Fig. (\ref{fig:5}). The Tracy-Widom distribution is obtained by using the R package called RMTstat \cite{Rmt14}. 

\begin{figure}
\includegraphics[width=0.5\textwidth]{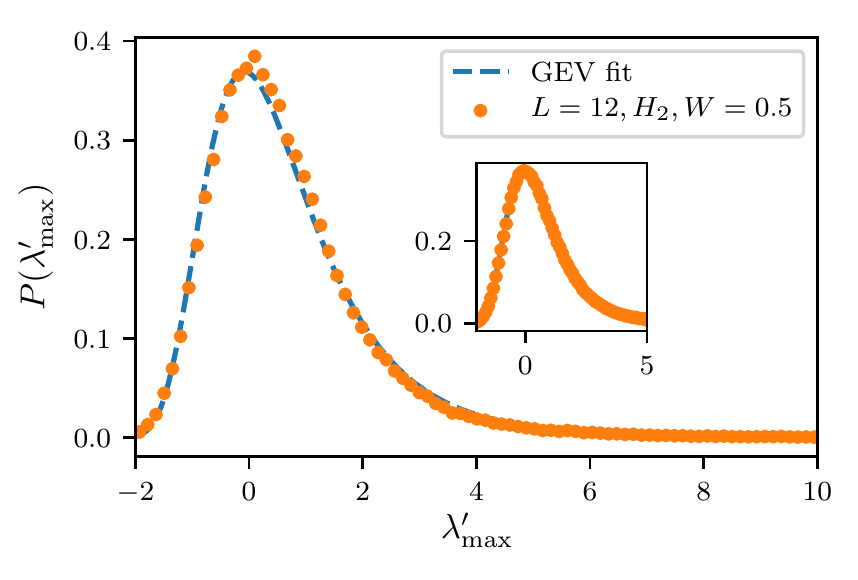}
\caption{Fit with the generalized extreme value distribution of the density of  $\lambda'_{\max}$ obtained from   Hamiltonian $H_2$ ($L=12$) and Hamiltonian $H_1$ ($L=14$,inset), $W=0.5$.}
\label{fig:6}
\end{figure}

As is clear, there are considerable deviations much beyond the finite size effects from the Tracy-Widom distribution, which implies that the correlations between eigenvalues of the reduced density matrix are not Wishart like. Thus, surprisingly even though the NNS distribution and number variance of the levels match GOE predictions for $H_1$ and $H_2$ (\cite{PH10,Bertrand16}), a more rigorous test with respect to extreme statistics shows that the correlations between eigenvalues 
of the  reduced density matrix of eigenvectors of the spin chains are much weaker. 
In  Fig. (\ref{fig:6}) we show a fit of the data 
with the generalized extreme value distribution with the probability density function,
\beq
\label{eq:gev-pdf}
\begin{split}
f(y;\xi) &=   \exp{(-(1-\xi y)^{1/\xi})} (1- \xi y)^{(1/\xi-1)} \; \mbox{for} \; \xi \ne 0  \\  
&=  \exp(-\exp(-y)) \exp(-y) \; \mbox{for} \; \xi=0. 
\end{split}.
\eeq

The shape parameter $\xi$ takes a value of $0.069$ and $-0.086$ respectively, for  $L=14$,  $H_1$ data and  $L=12$,  $H_2$ data for $W=0.5$. The data has been centered
and scaled by the location and scale parameters obtained by  fitting a generalized extreme value distribution with free location and 
scale parameters (using Scipy) so that 
Eq.~(\ref{eq:gev-pdf}) can be used. In all the plots involving GEV we 
plot $\lambda'_{\max}= (\lambda_1 - \mbox{loc})/\mbox{scale}$ with  the location (loc) and scale 
parameters for the fit produced in Table \ref{tab:table1}. Note, that as the location and scale parameters for fitting the Tracy-Widom distribution, given by
Eq.~(\ref{eq:scaledlambmax}) are different from that of the GEV distribution, $\lambda'_{\max}$ and $\lambda'_{1}$ are in general different. For the  $L=12,H_2$ data the location and scale parameters obtained from the fit are respectively
$0.069, 0.007$  while for the $L=14, H_1$ data they are respectively $0.053, 0.008$. 
This indicates that the best fitted distributions  are close to 
Fisher-Tipett-Gumbel.

\begin{figure}
\includegraphics[width=0.5\textwidth]{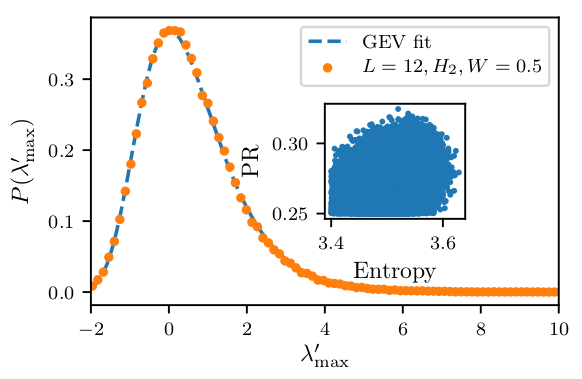} 
\caption{Fit with the generalized extreme value distribution of the density of  $\lambda'_{\max}$ obtained from  Hamiltonian $H_2$ for $W=0.5$, $L=12$, with an entropy and PR filter  of respectively $3.4$ and $0.25$. Inset shows the 
entropy vs. PR scatter plot.}
\label{fig:extx-filter}
\end{figure}

An interesting question is whether the deviation from the Wishart ensemble is due to the presence of low entropy and low PR states. To check for 
this we select eigenstates for the $H_2$ model, for $L=12$ with entropy and PR respectively greater than 
$3.4$ and $0.25$.  We then try to fit the maximum of the entanglement spectrum obtained from this data to the GEV distribution. The result is shown in Fig.~(\ref{fig:extx-filter}), and the $\xi$ value for the fit is $0.021$ (location and scale parameters for the fit respectively
being $ 0.073, 0.006$), indicating that the distribution
moves closer to being Fisher-Tipett-Gumbel, confirming that the correlations are not Wishart like even for the most random eigenstates.
The distribution of the maximum obtained from this filtered data also show a deviation from the Tracy-Widom distribution similar to the 
deviation shown in Fig.~(\ref{fig:5}), when we try to fit it using
Eq.~(\ref{eq:scaledlambmax}).

\section{Extreme value statistics away from the ergodic phase}

\subsection{Persistence of GEV for moderate disorder strengths}
\begin{figure}
\includegraphics[width=0.5\textwidth]{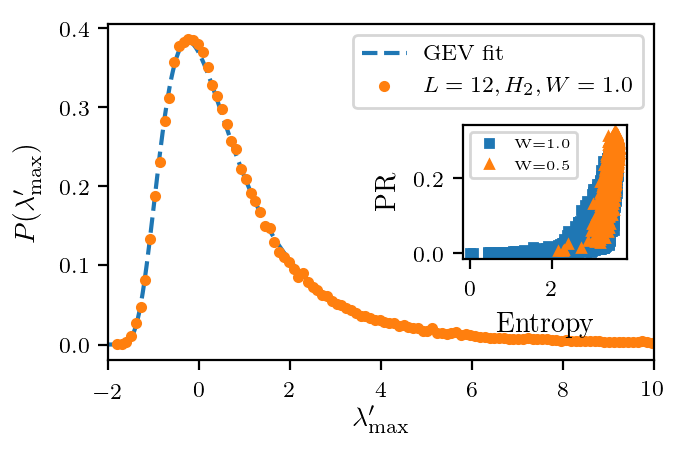} 
\caption{Fit with the generalized extreme value distribution of the density of  $\lambda'_{\max}$ obtained from  Hamiltonian 
$H_2$ for $W=1.0$, $L=12$. Inset shows Entropy {\it vs} PR for  $L=12$, the Hamiltonian $H_2$, $W=1.0$(squares) and $W=0.5$(triangles)}
\label{fig:w1}
\end{figure}

Figure~\ref{fig:w1} inset shows the entropy-PR scatter plot for the $H_2$ model for  disorder strength of $W=1.0$. The entropy and PR distribution spreads considerably with the mean entropy and mean PR respectively lowering to become $3.37$ and $0.17$ respectively.
The variance of entanglement entropy peaks as one approaches the transition, and thus the distributions are broadened. 
However, interestingly as shown in the main part of the same figure, the maximum still fits a GEV well, but with the $\xi$ parameter
being equal to $-0.3021$. 

\begin{figure}
\includegraphics[width=0.4\textwidth]{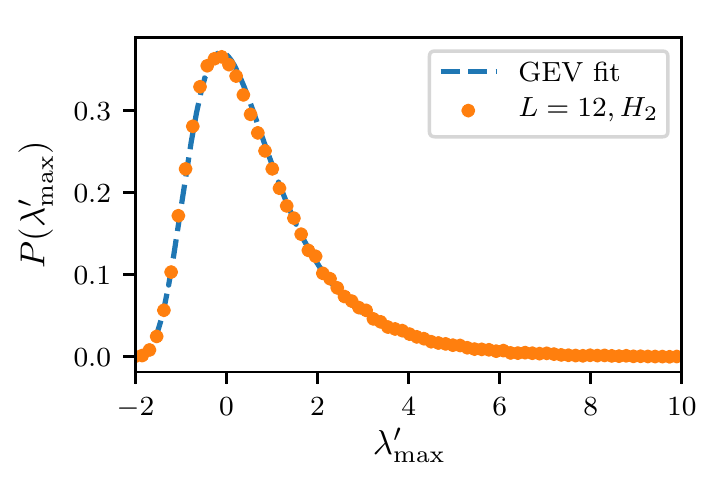} 
\caption{Fit with the generalized extreme value distribution of the density of  $\lambda'_{\max}$ obtained from  Hamiltonian $H_2$ for $W=1.0$ for eigenstates with entropy greater than $3.0$ and PR greater than $0.1$.}
\label{fig:w1filt}
\end{figure}

\begin{figure}
\includegraphics[width=0.5\textwidth]{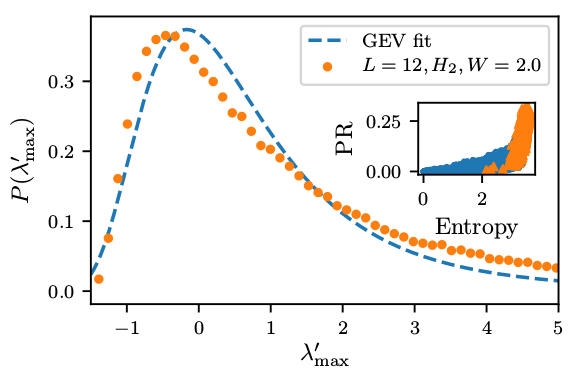} 
\caption{Fit with the generalized extreme value distribution of the density of  $\lambda'_{\max}$ obtained from  $H_2$ for $W=2.0$,  $L=12$. 
Entropy vs PR scatter plot shown in inset for  $L=12$, the Hamiltonian $H_2$, $W=2.0$ (dots) and 
$W=0.5$ ( triangles).}
\label{fig:w2}
\end{figure}

\begin{figure}
\includegraphics[width=0.4\textwidth]{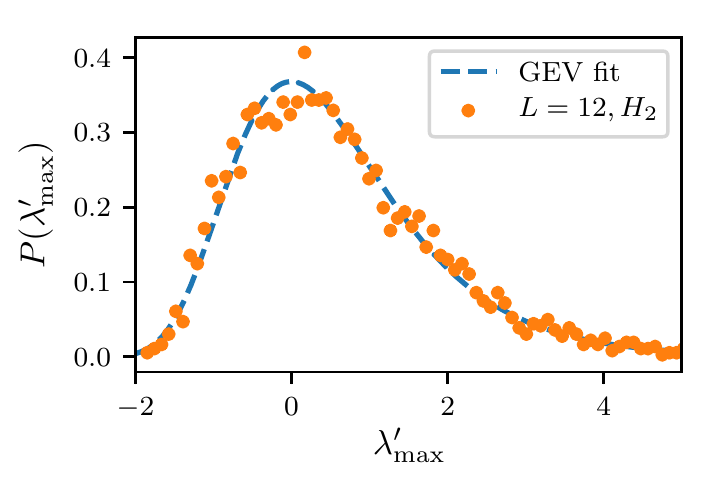} 
\caption{Fit with the generalized extreme value distribution of the density of  $\lambda'_{\max}$ obtained from  $H_2$ for $W=2.0$,  $L=12$ for eigenstates with entropy greater than $3.0$ and PR greater than $0.1$.}
\label{fig:w2filt}
\end{figure}

As we move more towards the localized phase in $W$, the eigenstates include many more low entropy and PR states indicating that they 
are almost localized and the maximum distribution seems to stop fitting a GEV. This is shown in Fig.~(\ref{fig:w2}) for $W=2.0$.
However, if we again weed out states with an entropy and PR respective lower than $3.0$ and $0.1$, the distribution again
moves close to a GEV with $\xi=-0.004$, thus close to the Fisher-Tipett-Gumbel distribution, see Fig.~\ref{fig:w2filt}. 

The different values of the shape parameter $\xi$, and location and scale parameters used for centering and scaling the data before
using Eq.~(\ref{eq:gev-pdf}) for the different fits for $L=12$, $H_2$ model  are collected in table \ref{tab:table1}. 
Beyond  about $W=2.0$ and closer to the transition the GEV distribution itself deviate much more significantly. 

\begin{table}[h!]
  \begin{center}
    \caption{$\xi$ value for GEV fit}
    \label{tab:table1}
    \begin{tabular}{|c|c|c|c|c|} 
    \hline
    \textbf{W} & \textbf{Filtered} &  \textbf{$\xi$ } & \textbf{loc} & \textbf{scale}  \\
      \hline
       $0.5$ & No  & $-0.086$ &  $0.069$  & $0.007$ \\                                   
       $0.5$ & Yes  & $-0.021$  &  $0.073$  & $0.006 $\\
       $1.0$ & No  & $-0.3021$ &  $0.092 $   &  $0.018$\\
       $1.0$ & Yes  & $-0.129$  &   $0.090 $ & $0.015 $ \\
       $2.0$ & Yes  & $-0.004$ &  $0.103 $   &  $0.021 $\\
      \hline
       \end{tabular}
  \end{center}
\end{table}

\subsection{Distribution of the maximum and second maximum of the entanglement spectrum and  power laws in the MBL phase}

\begin{figure}
\includegraphics[width=0.4\textwidth]{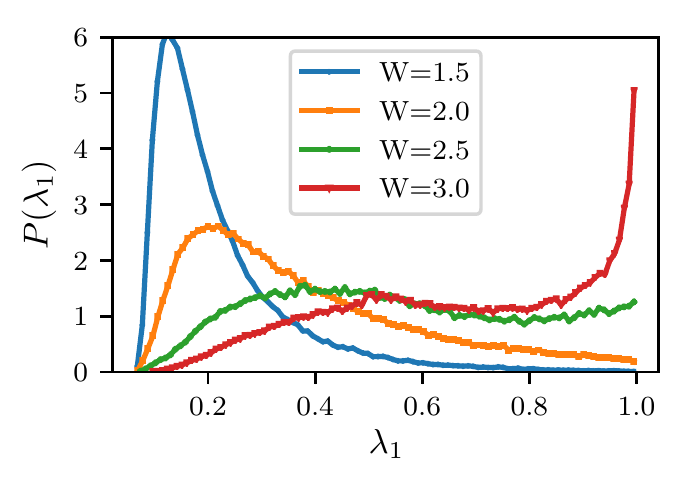} 
\includegraphics[width=0.4\textwidth]{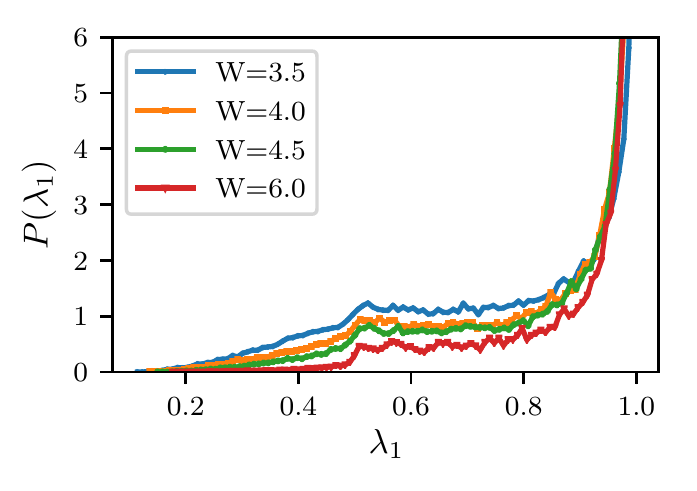} 
\caption{Distribution of the largest eigenvalue $\lambda_1$ for the Hamiltonian $H_2$, with $L=12$ as the disorder strength is increased across the MBL transition. }
\label{fig:maxdistrox}
\end{figure}
\begin{figure}
\includegraphics[width=0.4\textwidth]{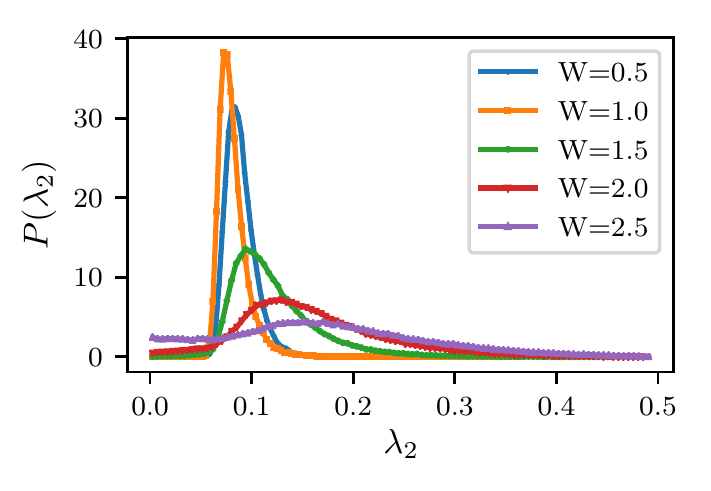} 
\includegraphics[width=0.4\textwidth]{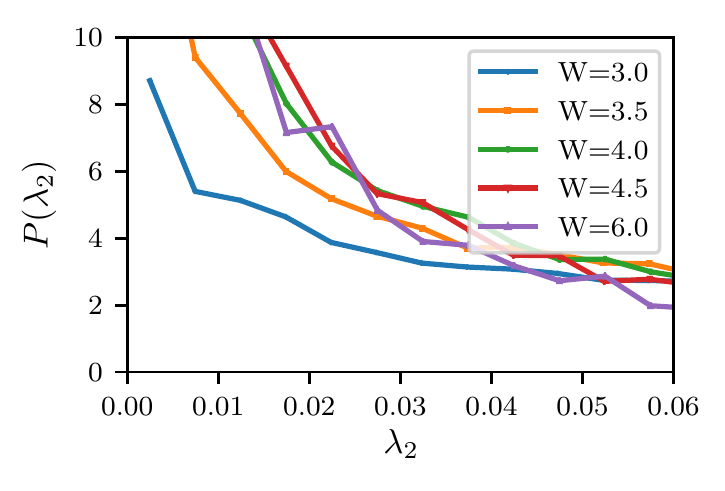} 
\caption{Distribution of $\lambda_2$ for $H_2$, $L=12$. }
\label{fig:secmaxdistrox}
\end{figure}

For the sake of compactness and simplicity we present results only for the model $H_2$ without total spin or particle number conservation, although we have verified the same for the $H_1$ case as well. Figures ~(\ref{fig:maxdistrox}) and (\ref{fig:secmaxdistrox}) show the distribution of unscaled or shifted (``raw") largest and second largest eigenvalue of the reduced density matrices, $\lambda_1$ and $\lambda_2$, for $W=1.5$ to $5$ for
the $H_2$ model.
While for $W=2.5$ the distribution of the 
largest eigenvalue $\lambda_1$ is rather broad, compared to the ergodic cases for $W=3.0$ it displays a peak around $\lambda_1=1$, indicating the extreme nature of the extreme, as the other eigenvalues are then forced to be of far lesser significance. The dominance of the largest eigenvalue is 
indicative of entry into MBL regimes. 
A feature, we mention in passing, in the distribution of the maximum is the kink that develops at $\lambda_1=1/2$ around $W=3.0$.
Such a feature has also been seen
in disorder averaged entanglement entropy as a resonance at $\ln{2}$ in \cite{Luitz16,limsheng16}, as well in weakly coupled chaotic systems \cite{Tomsovic2018}, originating in fact from the behavior of the dominating largest eigenvalue. 


%
%

\begin{figure}[!htb]
\includegraphics[width=0.4\textwidth]{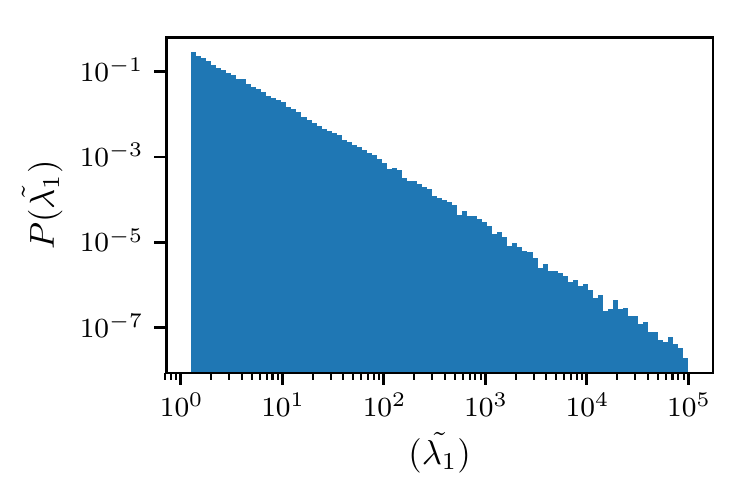} 
\caption{Distribution of $\tilde{\lambda_1}$ for the Hamiltonian $H_2$, and $L=12$, 
with the disorder strength $W=6$, the slope is $\approx$  $-1.45$ }
\label{fig:max1pow}
\end{figure}


\begin{figure}[!htb]
\includegraphics[width=0.4\textwidth]{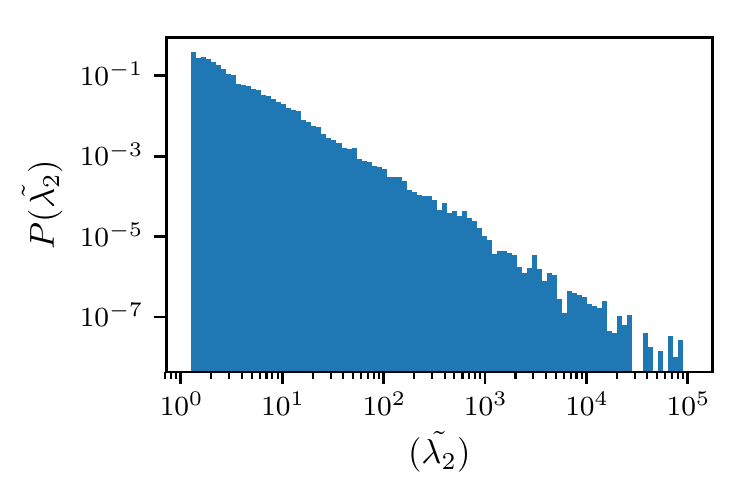} 
\caption{Distribution of $\tilde{\lambda_2}$ for the Hamiltonian $H_2$, and $L=12$, with the disorder strength $W=6$, the slope is $\approx$  $-1.7$.}
\label{fig:max2powx}
\end{figure}

As the disorder is increased the entanglement tends to the area-law and the largest eigenvalue tends to $1$. 
As for the ergodic phase, with increasing disorder, the GEV statistics seems to apply well only if we filter states that are sufficiently 
ergodic, we expect that in the transition regime and in the MBL phase itself it will be harder to control this and as the 
largest eigenvalue has become $O(1)$ rather than of $O(1/n)$, we shift our analysis away from the GEV framework. For large disorder strengths $W$, 
we may take the view that the system is one consisting of  spins with a Hamiltonian $\sum_{i}h_i' \sigma^z_i$ with $h_i'$ being uniformly distributed in
$[-1,1]$ being subjected to weak interactions of the order $1/W$ due to the other terms. Thus the unperturbed Hamiltonian with $W=\infty$
has a Poisson spectrum on which there is weak coupling that leads to resonances between the bare states. 

Recently a theory for such a scenario has been developed in \cite{Tomsovic2018} albeit in bipartite systems of weakly interacting chaotic systems.
Notably, the noninteracting case there leads to a Poisson spectrum (despite the chaos), and the largest and second largest 
eigenvalues of the reduced density matrices of eigenstates were heavy tailed, including the L\'evy distribution.
In the present case of many-body MBL, the partition is bipartite as well, while the individual subsystems in the non-interacting 
case are not chaotic, but show Poisson statistics. The key elements of the analysis rely more on the Poisson nature of the 
uncoupled systems and hence it is interesting to compare the extreme value results from there. We will ignore an overall scaling
by a ``transition parameter" that is yet to be identified, if at all it exists, in the case of the MBL transition.
Stable L\'evy laws were seen, as a consequence of combining a regularized perturbation theory with the generalized central limit theorem \cite{Tomsovic2018} in the densities of somewhat transformed quantities $\tilde{\lambda}_1=g(\lambda_1)$ and $\tilde{\lambda}_2=g(\lambda_2)$ 
where 
\beq
g(x)= \frac{x(1-x)}{(1-2x)^2}.
\eeq
Notice that $g(x)=g(1-x)$ and that for $\lambda_1$, as it is typically close to $1$ this is essentially $1-\lambda_1$, and for $\lambda_2$ which is typically $\ll 1$, this is $\approx \lambda_2$ itself. However, we are really interested in the excursions of these values into non-typical values which 
happens frequently due to the power-laws. Note also that $\lambda_2\leq 1/2$. Thus, applying the same transformation, Fig.~(\ref{fig:max1pow}) shows distribution for $\tilde{\lambda_1}$  for  the $H_2$ model for $W=6.0$, fairly deep in the MBL regime. An excellent power-law $\sim x^{-1.45}$ seems to 
be obtained, while Figure ~(\ref{fig:max2powx}) shows the distribution of $\tilde{\lambda_2}$, for the $H_2$ model and once again a power law tail is evident and is close to $\sim x^{-1.7}$.These are close indeed to the ones derived in \cite{Tomsovic2018} which is $\sim x^{-3/2}$ corresponding to both the cases.
Thus the extreme value statistics of MBL are also described by stable L\'evy laws that manifest clearly on an appropriate transformation. There is no doubt that the comparisons with the perturbation results derived in \cite{Tomsovic2018} need to be more critical, but we are  encouraged by the unmistakable similarities with the distribution of say $\lambda_1$ as it undergoes a transition into a localized regime, the broadening and the peak at $1/2$ are also seen in corresponding quantities of weakly coupled chaotic systems at appropriate coupling strengths which play the role of $W$.

\section{Conclusions}
In this work we have probed the randomness of the eigenstates of disordered spin chains showing the many-body localization transition, in the
ergodic phase. We have used extreme statistics of the entanglement spectrum for this purpose. This is much more sensitive to the randomness of 
eigenstates compared to the distribution of the average density of the entanglement spectra which follows nearly a Marchenko-Pastur distribution for
random states. For random states, the eigenvalues come from a trace constrained Wishart ensemble and the maximum follows the Tracy-Widom distribution after
a suitable shift of center and scaling. We have found significant deviations from the Tracy-Widom law and the distribution we obtained instead fits the 
Fisher-Tippet-Gumbel distribution quite well. This is the distribution that maximal events of independent and identically distributed random variables
follow after suitable rescaling. Our results thus indicate that even deep in the metallic phase the correlations in the entanglement spectra are not
as strong as those coming from truly random states and perhaps show signs of pre-localization. A natural question is if this is due to the presence of low 
entropy and participation ratio eigenstates present among the states which are closer to being random. But we have found that even the 
conditional distributions obtained after weeding out such states follow the generalized extreme value distribution and not the Tracy-Widom statistic. 
Even when the maximum distribution starts deviating significantly from GEV as we move towards the transition point, we found that the 
conditional distribution still follows GEV.

The fact that the Tracy-Widom law is not obtained even in the ergodic phase of a model with no conservation laws 
other than energy, maybe a generic feature of many-body systems. While quantities like the nearest 
neighbor spacing distributions or ratio of spacings may still be that of random matrices, the deviations 
will show in such properties of the eigenstates and the extreme value statistics could be the most stringent test of randomness, or 
at least one of them. It is possible though that further breaking conservation laws by Floquet kicked systems or even
non-periodically driven systems may restore the extreme values to be of the Tracy-Widom type. Excellent agreement with the
Marchenko-Pastur law and the Page value of subsystem entropy have been noted in these cases, for example see \cite{Mishra2014}. 

In the localized phase we shifted attention, motivated by recent results from a perturbation analysis of weakly coupled chaotic systems.
Remarkably, we found that suitably transforming the largest and second largest eigenvalues lead to power-law distributions that match reasonably well with that derived earlier and indicate the applicability of L\'evy stable laws in this context. This may also indicate the applicability of a regularized perturbation theory in the deep MBL regime, if not close to the transition. 
A work has appeared since the beginning of ours with identical motivations \cite{Gumbel19}. While our results unambiguously indicate the presence of the Fisher-Tippett-Gumbel law in the ergodic phase, the corresponding work shifts attention to the transition or localized regime. 
We have instead focused on an alternative strategy in the deep MBL phase wherein there are power-laws in the distribution of extreme eigenvalues
(quite distinct from power laws in disorder averaged entanglement spectra itself plotted against the eigenvalue order as found in \cite{Serbyn16}). However, while we found it untenable with our data to fit GEV distributions for large values of $W$, the results of  \cite{Gumbel19} indicate that this may still be possible.
Apart from a closer comparison with such works, ours would hopefully contribute to an understanding of extreme value statistics in the spectra of many-body systems. 

%
%
%
%

\bibliography{extmbl}

\begin{thebibliography}{56}%
\makeatletter
\providecommand \@ifxundefined [1]{%
 \@ifx{#1\undefined}
}%
\providecommand \@ifnum [1]{%
 \ifnum #1\expandafter \@firstoftwo
 \else \expandafter \@secondoftwo
 \fi
}%
\providecommand \@ifx [1]{%
 \ifx #1\expandafter \@firstoftwo
 \else \expandafter \@secondoftwo
 \fi
}%
\providecommand \natexlab [1]{#1}%
\providecommand \enquote  [1]{``#1''}%
\providecommand \bibnamefont  [1]{#1}%
\providecommand \bibfnamefont [1]{#1}%
\providecommand \citenamefont [1]{#1}%
\providecommand \href@noop [0]{\@secondoftwo}%
\providecommand \href [0]{\begingroup \@sanitize@url \@href}%
\providecommand \@href[1]{\@@startlink{#1}\@@href}%
\providecommand \@@href[1]{\endgroup#1\@@endlink}%
\providecommand \@sanitize@url [0]{\catcode `\\12\catcode `\$12\catcode
  `\&12\catcode `\#12\catcode `\^12\catcode `\_12\catcode `\%12\relax}%
\providecommand \@@startlink[1]{}%
\providecommand \@@endlink[0]{}%
\providecommand \url  [0]{\begingroup\@sanitize@url \@url }%
\providecommand \@url [1]{\endgroup\@href {#1}{\urlprefix }}%
\providecommand \urlprefix  [0]{URL }%
\providecommand \Eprint [0]{\href }%
\providecommand \doibase [0]{http://dx.doi.org/}%
\providecommand \selectlanguage [0]{\@gobble}%
\providecommand \bibinfo  [0]{\@secondoftwo}%
\providecommand \bibfield  [0]{\@secondoftwo}%
\providecommand \translation [1]{[#1]}%
\providecommand \BibitemOpen [0]{}%
\providecommand \bibitemStop [0]{}%
\providecommand \bibitemNoStop [0]{.\EOS\space}%
\providecommand \EOS [0]{\spacefactor3000\relax}%
\providecommand \BibitemShut  [1]{\csname bibitem#1\endcsname}%
\let\auto@bib@innerbib\@empty
\bibitem [{\citenamefont {Anderson}(1958)}]{And58}%
  \BibitemOpen
  \bibfield  {author} {\bibinfo {author} {\bibfnamefont {P.~W.}\ \bibnamefont
  {Anderson}},\ }\bibfield  {title} {\enquote {\bibinfo {title} {Absence of
  diffusion in certain random lattices},}\ }\href {\doibase
  10.1103/PhysRev.109.1492} {\bibfield  {journal} {\bibinfo  {journal} {Phys.
  Rev.}\ }\textbf {\bibinfo {volume} {109}},\ \bibinfo {pages} {1492--1505}
  (\bibinfo {year} {1958})}\BibitemShut {NoStop}%
\bibitem [{\citenamefont {Basko}\ \emph {et~al.}(2006)\citenamefont {Basko},
  \citenamefont {Aleiner},\ and\ \citenamefont {Altshuler}}]{Basko06}%
  \BibitemOpen
  \bibfield  {author} {\bibinfo {author} {\bibfnamefont {D.M.}\ \bibnamefont
  {Basko}}, \bibinfo {author} {\bibfnamefont {I.L.}\ \bibnamefont {Aleiner}}, \
  and\ \bibinfo {author} {\bibfnamefont {B.L.}\ \bibnamefont {Altshuler}},\
  }\bibfield  {title} {\enquote {\bibinfo {title} {Metal–insulator transition
  in a weakly interacting many-electron system with localized single-particle
  states},}\ }\href {\doibase https://doi.org/10.1016/j.aop.2005.11.014}
  {\bibfield  {journal} {\bibinfo  {journal} {Annals of Physics}\ }\textbf
  {\bibinfo {volume} {321}},\ \bibinfo {pages} {1126 -- 1205} (\bibinfo {year}
  {2006})}\BibitemShut {NoStop}%
\bibitem [{\citenamefont {Imbrie}(2016)}]{Imbrie16}%
  \BibitemOpen
  \bibfield  {author} {\bibinfo {author} {\bibfnamefont {John~Z.}\ \bibnamefont
  {Imbrie}},\ }\bibfield  {title} {\enquote {\bibinfo {title} {Diagonalization
  and many-body localization for a disordered quantum spin chain},}\ }\href
  {\doibase 10.1103/PhysRevLett.117.027201} {\bibfield  {journal} {\bibinfo
  {journal} {Phys. Rev. Lett.}\ }\textbf {\bibinfo {volume} {117}},\ \bibinfo
  {pages} {027201} (\bibinfo {year} {2016})}\BibitemShut {NoStop}%
\bibitem [{\citenamefont {Nandkishore}\ and\ \citenamefont
  {Huse}(2015)}]{HN15}%
  \BibitemOpen
  \bibfield  {author} {\bibinfo {author} {\bibfnamefont {Rahul}\ \bibnamefont
  {Nandkishore}}\ and\ \bibinfo {author} {\bibfnamefont {David~A.}\
  \bibnamefont {Huse}},\ }\bibfield  {title} {\enquote {\bibinfo {title}
  {Many-body localization and thermalization in quantum statistical
  mechanics},}\ }\href {\doibase 10.1146/annurev-conmatphys-031214-014726}
  {\bibfield  {journal} {\bibinfo  {journal} {Annual Review of Condensed Matter
  Physics}\ }\textbf {\bibinfo {volume} {6}},\ \bibinfo {pages} {15--38}
  (\bibinfo {year} {2015})},\ \Eprint
  {http://arxiv.org/abs/https://doi.org/10.1146/annurev-conmatphys-031214-014726}
  {https://doi.org/10.1146/annurev-conmatphys-031214-014726} \BibitemShut
  {NoStop}%
\bibitem [{\citenamefont {Huse}\ \emph {et~al.}(2014)\citenamefont {Huse},
  \citenamefont {Nandkishore},\ and\ \citenamefont {Oganesyan}}]{Huse14}%
  \BibitemOpen
  \bibfield  {author} {\bibinfo {author} {\bibfnamefont {David~A.}\
  \bibnamefont {Huse}}, \bibinfo {author} {\bibfnamefont {Rahul}\ \bibnamefont
  {Nandkishore}}, \ and\ \bibinfo {author} {\bibfnamefont {Vadim}\ \bibnamefont
  {Oganesyan}},\ }\bibfield  {title} {\enquote {\bibinfo {title} {Phenomenology
  of fully many-body-localized systems},}\ }\href {\doibase
  10.1103/PhysRevB.90.174202} {\bibfield  {journal} {\bibinfo  {journal} {Phys.
  Rev. B}\ }\textbf {\bibinfo {volume} {90}},\ \bibinfo {pages} {174202}
  (\bibinfo {year} {2014})}\BibitemShut {NoStop}%
\bibitem [{\citenamefont {Bardarson}\ \emph
  {et~al.}(2012{\natexlab{a}})\citenamefont {Bardarson}, \citenamefont
  {Pollmann},\ and\ \citenamefont {Moore}}]{BPM12}%
  \BibitemOpen
  \bibfield  {author} {\bibinfo {author} {\bibfnamefont {Jens~H.}\ \bibnamefont
  {Bardarson}}, \bibinfo {author} {\bibfnamefont {Frank}\ \bibnamefont
  {Pollmann}}, \ and\ \bibinfo {author} {\bibfnamefont {Joel~E.}\ \bibnamefont
  {Moore}},\ }\bibfield  {title} {\enquote {\bibinfo {title} {Unbounded growth
  of entanglement in models of many-body localization},}\ }\href {\doibase
  10.1103/PhysRevLett.109.017202} {\bibfield  {journal} {\bibinfo  {journal}
  {Phys. Rev. Lett.}\ }\textbf {\bibinfo {volume} {109}},\ \bibinfo {pages}
  {017202} (\bibinfo {year} {2012}{\natexlab{a}})}\BibitemShut {NoStop}%
\bibitem [{\citenamefont {Pal}\ and\ \citenamefont {Huse}(2010)}]{PH10}%
  \BibitemOpen
  \bibfield  {author} {\bibinfo {author} {\bibfnamefont {Arijeet}\ \bibnamefont
  {Pal}}\ and\ \bibinfo {author} {\bibfnamefont {David~A.}\ \bibnamefont
  {Huse}},\ }\bibfield  {title} {\enquote {\bibinfo {title} {Many-body
  localization phase transition},}\ }\href {\doibase
  10.1103/PhysRevB.82.174411} {\bibfield  {journal} {\bibinfo  {journal} {Phys.
  Rev. B}\ }\textbf {\bibinfo {volume} {82}},\ \bibinfo {pages} {174411}
  (\bibinfo {year} {2010})}\BibitemShut {NoStop}%
\bibitem [{\citenamefont {Yao~N.Y}\ and\ \citenamefont {A.}(2015)}]{YLV15}%
  \BibitemOpen
  \bibfield  {author} {\bibinfo {author} {\bibfnamefont {Laumann~C.R}\
  \bibnamefont {Yao~N.Y}}\ and\ \bibinfo {author} {\bibfnamefont {Vishwanath}\
  \bibnamefont {A.}},\ }\bibfield  {title} {\enquote {\bibinfo {title}
  {Many-body localization protected quantum state transfer},}\ }\href@noop {}
  {\bibfield  {journal} {\bibinfo  {journal} {arXiv:1508.06995}\ } (\bibinfo
  {year} {2015})}\BibitemShut {NoStop}%
\bibitem [{\citenamefont {Serbyn}\ \emph {et~al.}(2014)\citenamefont {Serbyn},
  \citenamefont {Knap}, \citenamefont {Gopalakrishnan}, \citenamefont
  {Papi\ifmmode~\acute{c}\else \'{c}\fi{}}, \citenamefont {Yao}, \citenamefont
  {Laumann}, \citenamefont {Abanin}, \citenamefont {Lukin},\ and\ \citenamefont
  {Demler}}]{Demlu14}%
  \BibitemOpen
  \bibfield  {author} {\bibinfo {author} {\bibfnamefont {M.}~\bibnamefont
  {Serbyn}}, \bibinfo {author} {\bibfnamefont {M.}~\bibnamefont {Knap}},
  \bibinfo {author} {\bibfnamefont {S.}~\bibnamefont {Gopalakrishnan}},
  \bibinfo {author} {\bibfnamefont {Z.}~\bibnamefont
  {Papi\ifmmode~\acute{c}\else \'{c}\fi{}}}, \bibinfo {author} {\bibfnamefont
  {N.~Y.}\ \bibnamefont {Yao}}, \bibinfo {author} {\bibfnamefont {C.~R.}\
  \bibnamefont {Laumann}}, \bibinfo {author} {\bibfnamefont {D.~A.}\
  \bibnamefont {Abanin}}, \bibinfo {author} {\bibfnamefont {M.~D.}\
  \bibnamefont {Lukin}}, \ and\ \bibinfo {author} {\bibfnamefont {E.~A.}\
  \bibnamefont {Demler}},\ }\bibfield  {title} {\enquote {\bibinfo {title}
  {Interferometric probes of many-body localization},}\ }\href {\doibase
  10.1103/PhysRevLett.113.147204} {\bibfield  {journal} {\bibinfo  {journal}
  {Phys. Rev. Lett.}\ }\textbf {\bibinfo {volume} {113}},\ \bibinfo {pages}
  {147204} (\bibinfo {year} {2014})}\BibitemShut {NoStop}%
\bibitem [{\citenamefont {Schreiber}\ \emph {et~al.}(2015)\citenamefont
  {Schreiber}, \citenamefont {Hodgman}, \citenamefont {Bordia}, \citenamefont
  {L{\"u}schen}, \citenamefont {Fischer}, \citenamefont {Vosk}, \citenamefont
  {Altman}, \citenamefont {Schneider},\ and\ \citenamefont
  {Bloch}}]{Schreiber15}%
  \BibitemOpen
  \bibfield  {author} {\bibinfo {author} {\bibfnamefont {Michael}\ \bibnamefont
  {Schreiber}}, \bibinfo {author} {\bibfnamefont {Sean~S.}\ \bibnamefont
  {Hodgman}}, \bibinfo {author} {\bibfnamefont {Pranjal}\ \bibnamefont
  {Bordia}}, \bibinfo {author} {\bibfnamefont {Henrik~P.}\ \bibnamefont
  {L{\"u}schen}}, \bibinfo {author} {\bibfnamefont {Mark~H.}\ \bibnamefont
  {Fischer}}, \bibinfo {author} {\bibfnamefont {Ronen}\ \bibnamefont {Vosk}},
  \bibinfo {author} {\bibfnamefont {Ehud}\ \bibnamefont {Altman}}, \bibinfo
  {author} {\bibfnamefont {Ulrich}\ \bibnamefont {Schneider}}, \ and\ \bibinfo
  {author} {\bibfnamefont {Immanuel}\ \bibnamefont {Bloch}},\ }\bibfield
  {title} {\enquote {\bibinfo {title} {Observation of many-body localization of
  interacting fermions in a quasirandom optical lattice},}\ }\href {\doibase
  10.1126/science.aaa7432} {\bibfield  {journal} {\bibinfo  {journal}
  {Science}\ }\textbf {\bibinfo {volume} {349}},\ \bibinfo {pages} {842--845}
  (\bibinfo {year} {2015})},\ \Eprint
  {http://arxiv.org/abs/http://science.sciencemag.org/content/349/6250/842.full.pdf}
  {http://science.sciencemag.org/content/349/6250/842.full.pdf} \BibitemShut
  {NoStop}%
\bibitem [{\citenamefont {Amico}\ \emph {et~al.}(2008)\citenamefont {Amico},
  \citenamefont {Fazio}, \citenamefont {Osterloh},\ and\ \citenamefont
  {Vedral}}]{Amico08}%
  \BibitemOpen
  \bibfield  {author} {\bibinfo {author} {\bibfnamefont {Luigi}\ \bibnamefont
  {Amico}}, \bibinfo {author} {\bibfnamefont {Rosario}\ \bibnamefont {Fazio}},
  \bibinfo {author} {\bibfnamefont {Andreas}\ \bibnamefont {Osterloh}}, \ and\
  \bibinfo {author} {\bibfnamefont {Vlatko}\ \bibnamefont {Vedral}},\
  }\bibfield  {title} {\enquote {\bibinfo {title} {Entanglement in many-body
  systems},}\ }\href {\doibase 10.1103/RevModPhys.80.517} {\bibfield  {journal}
  {\bibinfo  {journal} {Rev. Mod. Phys.}\ }\textbf {\bibinfo {volume} {80}},\
  \bibinfo {pages} {517--576} (\bibinfo {year} {2008})}\BibitemShut {NoStop}%
\bibitem [{\citenamefont {Eisert}\ \emph {et~al.}(2010)\citenamefont {Eisert},
  \citenamefont {Cramer},\ and\ \citenamefont {Plenio}}]{Eisert10}%
  \BibitemOpen
  \bibfield  {author} {\bibinfo {author} {\bibfnamefont {J.}~\bibnamefont
  {Eisert}}, \bibinfo {author} {\bibfnamefont {M.}~\bibnamefont {Cramer}}, \
  and\ \bibinfo {author} {\bibfnamefont {M.~B.}\ \bibnamefont {Plenio}},\
  }\bibfield  {title} {\enquote {\bibinfo {title} {Colloquium: Area laws for
  the entanglement entropy},}\ }\href {\doibase 10.1103/RevModPhys.82.277}
  {\bibfield  {journal} {\bibinfo  {journal} {Rev. Mod. Phys.}\ }\textbf
  {\bibinfo {volume} {82}},\ \bibinfo {pages} {277--306} (\bibinfo {year}
  {2010})}\BibitemShut {NoStop}%
\bibitem [{\citenamefont {Kitaev}\ and\ \citenamefont
  {Preskill}(2006)}]{Kitaev06}%
  \BibitemOpen
  \bibfield  {author} {\bibinfo {author} {\bibfnamefont {Alexei}\ \bibnamefont
  {Kitaev}}\ and\ \bibinfo {author} {\bibfnamefont {John}\ \bibnamefont
  {Preskill}},\ }\bibfield  {title} {\enquote {\bibinfo {title} {Topological
  entanglement entropy},}\ }\href {\doibase 10.1103/PhysRevLett.96.110404}
  {\bibfield  {journal} {\bibinfo  {journal} {Phys. Rev. Lett.}\ }\textbf
  {\bibinfo {volume} {96}},\ \bibinfo {pages} {110404} (\bibinfo {year}
  {2006})}\BibitemShut {NoStop}%
\bibitem [{\citenamefont {Wen}(2003)}]{Wen03}%
  \BibitemOpen
  \bibfield  {author} {\bibinfo {author} {\bibfnamefont {Xiao-Gang}\
  \bibnamefont {Wen}},\ }\bibfield  {title} {\enquote {\bibinfo {title}
  {Quantum orders in an exact soluble model},}\ }\href {\doibase
  10.1103/PhysRevLett.90.016803} {\bibfield  {journal} {\bibinfo  {journal}
  {Phys. Rev. Lett.}\ }\textbf {\bibinfo {volume} {90}},\ \bibinfo {pages}
  {016803} (\bibinfo {year} {2003})}\BibitemShut {NoStop}%
\bibitem [{\citenamefont {Bardarson}\ \emph
  {et~al.}(2012{\natexlab{b}})\citenamefont {Bardarson}, \citenamefont
  {Pollmann},\ and\ \citenamefont {Moore}}]{Pollman12}%
  \BibitemOpen
  \bibfield  {author} {\bibinfo {author} {\bibfnamefont {Jens~H.}\ \bibnamefont
  {Bardarson}}, \bibinfo {author} {\bibfnamefont {Frank}\ \bibnamefont
  {Pollmann}}, \ and\ \bibinfo {author} {\bibfnamefont {Joel~E.}\ \bibnamefont
  {Moore}},\ }\bibfield  {title} {\enquote {\bibinfo {title} {Unbounded growth
  of entanglement in models of many-body localization},}\ }\href {\doibase
  10.1103/PhysRevLett.109.017202} {\bibfield  {journal} {\bibinfo  {journal}
  {Phys. Rev. Lett.}\ }\textbf {\bibinfo {volume} {109}},\ \bibinfo {pages}
  {017202} (\bibinfo {year} {2012}{\natexlab{b}})}\BibitemShut {NoStop}%
\bibitem [{\citenamefont {Kj\"all}\ \emph {et~al.}(2014)\citenamefont
  {Kj\"all}, \citenamefont {Bardarson},\ and\ \citenamefont
  {Pollmann}}]{Pollman14}%
  \BibitemOpen
  \bibfield  {author} {\bibinfo {author} {\bibfnamefont {Jonas~A.}\
  \bibnamefont {Kj\"all}}, \bibinfo {author} {\bibfnamefont {Jens~H.}\
  \bibnamefont {Bardarson}}, \ and\ \bibinfo {author} {\bibfnamefont {Frank}\
  \bibnamefont {Pollmann}},\ }\bibfield  {title} {\enquote {\bibinfo {title}
  {Many-body localization in a disordered quantum ising chain},}\ }\href
  {\doibase 10.1103/PhysRevLett.113.107204} {\bibfield  {journal} {\bibinfo
  {journal} {Phys. Rev. Lett.}\ }\textbf {\bibinfo {volume} {113}},\ \bibinfo
  {pages} {107204} (\bibinfo {year} {2014})}\BibitemShut {NoStop}%
\bibitem [{\citenamefont {Bera}\ and\ \citenamefont
  {Lakshminarayan}(2016)}]{Bera2016}%
  \BibitemOpen
  \bibfield  {author} {\bibinfo {author} {\bibfnamefont {Soumya}\ \bibnamefont
  {Bera}}\ and\ \bibinfo {author} {\bibfnamefont {Arul}\ \bibnamefont
  {Lakshminarayan}},\ }\bibfield  {title} {\enquote {\bibinfo {title} {Local
  entanglement structure across a many-body localization transition},}\ }\href
  {\doibase 10.1103/PhysRevB.93.134204} {\bibfield  {journal} {\bibinfo
  {journal} {Phys. Rev. B}\ }\textbf {\bibinfo {volume} {93}},\ \bibinfo
  {pages} {134204} (\bibinfo {year} {2016})}\BibitemShut {NoStop}%
\bibitem [{\citenamefont {Li}\ and\ \citenamefont {Haldane}(2008)}]{Haldane08}%
  \BibitemOpen
  \bibfield  {author} {\bibinfo {author} {\bibfnamefont {Hui}\ \bibnamefont
  {Li}}\ and\ \bibinfo {author} {\bibfnamefont {F.~D.~M.}\ \bibnamefont
  {Haldane}},\ }\bibfield  {title} {\enquote {\bibinfo {title} {Entanglement
  spectrum as a generalization of entanglement entropy: Identification of
  topological order in non-abelian fractional quantum hall effect states},}\
  }\href {\doibase 10.1103/PhysRevLett.101.010504} {\bibfield  {journal}
  {\bibinfo  {journal} {Phys. Rev. Lett.}\ }\textbf {\bibinfo {volume} {101}},\
  \bibinfo {pages} {010504} (\bibinfo {year} {2008})}\BibitemShut {NoStop}%
\bibitem [{\citenamefont {Yang}\ \emph {et~al.}(2015)\citenamefont {Yang},
  \citenamefont {Chamon}, \citenamefont {Hamma},\ and\ \citenamefont
  {Mucciolo}}]{Chamon15}%
  \BibitemOpen
  \bibfield  {author} {\bibinfo {author} {\bibfnamefont {Zhi-Cheng}\
  \bibnamefont {Yang}}, \bibinfo {author} {\bibfnamefont {Claudio}\
  \bibnamefont {Chamon}}, \bibinfo {author} {\bibfnamefont {Alioscia}\
  \bibnamefont {Hamma}}, \ and\ \bibinfo {author} {\bibfnamefont {Eduardo~R.}\
  \bibnamefont {Mucciolo}},\ }\bibfield  {title} {\enquote {\bibinfo {title}
  {Two-component structure in the entanglement spectrum of highly excited
  states},}\ }\href {\doibase 10.1103/PhysRevLett.115.267206} {\bibfield
  {journal} {\bibinfo  {journal} {Phys. Rev. Lett.}\ }\textbf {\bibinfo
  {volume} {115}},\ \bibinfo {pages} {267206} (\bibinfo {year}
  {2015})}\BibitemShut {NoStop}%
\bibitem [{\citenamefont {Geraedts}\ \emph {et~al.}(2016)\citenamefont
  {Geraedts}, \citenamefont {Nandkishore},\ and\ \citenamefont
  {Regnault}}]{Nand16}%
  \BibitemOpen
  \bibfield  {author} {\bibinfo {author} {\bibfnamefont {Scott~D.}\
  \bibnamefont {Geraedts}}, \bibinfo {author} {\bibfnamefont {Rahul}\
  \bibnamefont {Nandkishore}}, \ and\ \bibinfo {author} {\bibfnamefont
  {Nicolas}\ \bibnamefont {Regnault}},\ }\bibfield  {title} {\enquote {\bibinfo
  {title} {Many-body localization and thermalization: Insights from the
  entanglement spectrum},}\ }\href {\doibase 10.1103/PhysRevB.93.174202}
  {\bibfield  {journal} {\bibinfo  {journal} {Phys. Rev. B}\ }\textbf {\bibinfo
  {volume} {93}},\ \bibinfo {pages} {174202} (\bibinfo {year}
  {2016})}\BibitemShut {NoStop}%
\bibitem [{\citenamefont {Serbyn}\ \emph {et~al.}(2016)\citenamefont {Serbyn},
  \citenamefont {Michailidis}, \citenamefont {Abanin},\ and\ \citenamefont
  {Papi\ifmmode~\acute{c}\else \'{c}\fi{}}}]{Serbyn16}%
  \BibitemOpen
  \bibfield  {author} {\bibinfo {author} {\bibfnamefont {Maksym}\ \bibnamefont
  {Serbyn}}, \bibinfo {author} {\bibfnamefont {Alexios~A.}\ \bibnamefont
  {Michailidis}}, \bibinfo {author} {\bibfnamefont {Dmitry~A.}\ \bibnamefont
  {Abanin}}, \ and\ \bibinfo {author} {\bibfnamefont {Z.}~\bibnamefont
  {Papi\ifmmode~\acute{c}\else \'{c}\fi{}}},\ }\bibfield  {title} {\enquote
  {\bibinfo {title} {Power-law entanglement spectrum in many-body localized
  phases},}\ }\href {\doibase 10.1103/PhysRevLett.117.160601} {\bibfield
  {journal} {\bibinfo  {journal} {Phys. Rev. Lett.}\ }\textbf {\bibinfo
  {volume} {117}},\ \bibinfo {pages} {160601} (\bibinfo {year}
  {2016})}\BibitemShut {NoStop}%
\bibitem [{\citenamefont {Pietracaprina}\ \emph {et~al.}(2017)\citenamefont
  {Pietracaprina}, \citenamefont {Parisi}, \citenamefont {Mariano},
  \citenamefont {Pascazio},\ and\ \citenamefont {Scardicchio}}]{Parisi17}%
  \BibitemOpen
  \bibfield  {author} {\bibinfo {author} {\bibfnamefont {Francesca}\
  \bibnamefont {Pietracaprina}}, \bibinfo {author} {\bibfnamefont {Giorgio}\
  \bibnamefont {Parisi}}, \bibinfo {author} {\bibfnamefont {Angelo}\
  \bibnamefont {Mariano}}, \bibinfo {author} {\bibfnamefont {Saverio}\
  \bibnamefont {Pascazio}}, \ and\ \bibinfo {author} {\bibfnamefont
  {Antonello}\ \bibnamefont {Scardicchio}},\ }\bibfield  {title} {\enquote
  {\bibinfo {title} {Entanglement critical length at the many-body localization
  transition},}\ }\href {http://stacks.iop.org/1742-5468/2017/i=11/a=113102}
  {\bibfield  {journal} {\bibinfo  {journal} {Journal of Statistical Mechanics:
  Theory and Experiment}\ }\textbf {\bibinfo {volume} {2017}},\ \bibinfo
  {pages} {113102} (\bibinfo {year} {2017})}\BibitemShut {NoStop}%
\bibitem [{\citenamefont {Atas}\ \emph {et~al.}(2013)\citenamefont {Atas},
  \citenamefont {Bogomolny}, \citenamefont {Giraud},\ and\ \citenamefont
  {Roux}}]{Atas2013}%
  \BibitemOpen
  \bibfield  {author} {\bibinfo {author} {\bibfnamefont {Y.~Y.}\ \bibnamefont
  {Atas}}, \bibinfo {author} {\bibfnamefont {E.}~\bibnamefont {Bogomolny}},
  \bibinfo {author} {\bibfnamefont {O.}~\bibnamefont {Giraud}}, \ and\ \bibinfo
  {author} {\bibfnamefont {G.}~\bibnamefont {Roux}},\ }\bibfield  {title}
  {\enquote {\bibinfo {title} {Distribution of the ratio of consecutive level
  spacings in random matrix ensembles},}\ }\href {\doibase
  10.1103/PhysRevLett.110.084101} {\bibfield  {journal} {\bibinfo  {journal}
  {Phys. Rev. Lett.}\ }\textbf {\bibinfo {volume} {110}},\ \bibinfo {pages}
  {084101} (\bibinfo {year} {2013})}\BibitemShut {NoStop}%
\bibitem [{\citenamefont {Mehta}(2004)}]{Mehta}%
  \BibitemOpen
  \bibfield  {author} {\bibinfo {author} {\bibfnamefont {Madan~Lal}\
  \bibnamefont {Mehta}},\ }\href@noop {} {\emph {\bibinfo {title} {Random
  Matrices}}},\ \bibinfo {edition} {3rd}\ ed.\ (\bibinfo {year}
  {2004})\BibitemShut {NoStop}%
\bibitem [{\citenamefont {Giannoni}\ \emph {et~al.}(1991)\citenamefont
  {Giannoni}, \citenamefont {Voros},\ and\ \citenamefont
  {Zinn-Justin}}]{Leshousches89}%
  \BibitemOpen
  \bibfield  {author} {\bibinfo {author} {\bibfnamefont {Marie-Joya}\
  \bibnamefont {Giannoni}}, \bibinfo {author} {\bibfnamefont {Andre}\
  \bibnamefont {Voros}}, \ and\ \bibinfo {author} {\bibfnamefont {Jean}\
  \bibnamefont {Zinn-Justin}},\ }\href@noop {} {\emph {\bibinfo {title} {Chaos
  and quantum physics,Volume 52 of Les Houches Summer School Proceedings
  Series}}}\ (\bibinfo  {publisher} {North-Holland},\ \bibinfo {year}
  {1991})\BibitemShut {NoStop}%
\bibitem [{\citenamefont {Bertrand}\ and\ \citenamefont
  {Garc\'{\i}a-Garc\'{\i}a}(2016)}]{Bertrand16}%
  \BibitemOpen
  \bibfield  {author} {\bibinfo {author} {\bibfnamefont {Corentin~L.}\
  \bibnamefont {Bertrand}}\ and\ \bibinfo {author} {\bibfnamefont {Antonio~M.}\
  \bibnamefont {Garc\'{\i}a-Garc\'{\i}a}},\ }\bibfield  {title} {\enquote
  {\bibinfo {title} {Anomalous thouless energy and critical statistics on the
  metallic side of the many-body localization transition},}\ }\href {\doibase
  10.1103/PhysRevB.94.144201} {\bibfield  {journal} {\bibinfo  {journal} {Phys.
  Rev. B}\ }\textbf {\bibinfo {volume} {94}},\ \bibinfo {pages} {144201}
  (\bibinfo {year} {2016})}\BibitemShut {NoStop}%
\bibitem [{\citenamefont {Page}(1993)}]{Page93}%
  \BibitemOpen
  \bibfield  {author} {\bibinfo {author} {\bibfnamefont {Don~N.}\ \bibnamefont
  {Page}},\ }\bibfield  {title} {\enquote {\bibinfo {title} {Average entropy of
  a subsystem},}\ }\href {\doibase 10.1103/PhysRevLett.71.1291} {\bibfield
  {journal} {\bibinfo  {journal} {Phys. Rev. Lett.}\ }\textbf {\bibinfo
  {volume} {71}},\ \bibinfo {pages} {1291--1294} (\bibinfo {year}
  {1993})}\BibitemShut {NoStop}%
\bibitem [{\citenamefont {Tracy}\ and\ \citenamefont {Widom}(1996)}]{Tracy96}%
  \BibitemOpen
  \bibfield  {author} {\bibinfo {author} {\bibfnamefont {Craig~A.}\
  \bibnamefont {Tracy}}\ and\ \bibinfo {author} {\bibfnamefont {Harold}\
  \bibnamefont {Widom}},\ }\bibfield  {title} {\enquote {\bibinfo {title} {On
  orthogonal and symplectic matrix ensembles},}\ }\href {\doibase
  10.1007/BF02099545} {\bibfield  {journal} {\bibinfo  {journal}
  {Communications in Mathematical Physics}\ }\textbf {\bibinfo {volume}
  {177}},\ \bibinfo {pages} {727--754} (\bibinfo {year} {1996})}\BibitemShut
  {NoStop}%
\bibitem [{\citenamefont {Fisher}\ and\ \citenamefont
  {Tippett}(1928)}]{FisherTippett1928}%
  \BibitemOpen
  \bibfield  {author} {\bibinfo {author} {\bibfnamefont {R.~A.}\ \bibnamefont
  {Fisher}}\ and\ \bibinfo {author} {\bibfnamefont {L.~H.~C.}\ \bibnamefont
  {Tippett}},\ }\bibfield  {title} {\enquote {\bibinfo {title} {Limiting forms
  of the frequency distribution of the largest or smallest member of a
  sample},}\ }\href@noop {} {\bibfield  {journal} {\bibinfo  {journal}
  {Mathematical Proceedings of the Cambridge Philosophical Society}\ }\textbf
  {\bibinfo {volume} {24}},\ \bibinfo {pages} {180} (\bibinfo {year}
  {1928})}\BibitemShut {NoStop}%
\bibitem [{\citenamefont {Gumbel}(2004)}]{Gumbel04}%
  \BibitemOpen
  \bibfield  {author} {\bibinfo {author} {\bibfnamefont {E~J}\ \bibnamefont
  {Gumbel}},\ }\href@noop {} {\emph {\bibinfo {title} {Statistics of
  extremes}}}\ (\bibinfo  {publisher} {Dover PublicationsInc., New York},\
  \bibinfo {year} {2004})\BibitemShut {NoStop}%
\bibitem [{\citenamefont {Leadbetter}\ and\ \citenamefont
  {Rootzen}(1988)}]{Leadbetter88}%
  \BibitemOpen
  \bibfield  {author} {\bibinfo {author} {\bibfnamefont {M.~R.}\ \bibnamefont
  {Leadbetter}}\ and\ \bibinfo {author} {\bibfnamefont {Holger}\ \bibnamefont
  {Rootzen}},\ }\bibfield  {title} {\enquote {\bibinfo {title} {Extremal theory
  for stochastic processes},}\ }\href {http://www.jstor.org/stable/2243819}
  {\bibfield  {journal} {\bibinfo  {journal} {The Annals of Probability}\
  }\textbf {\bibinfo {volume} {16}},\ \bibinfo {pages} {431--478} (\bibinfo
  {year} {1988})}\BibitemShut {NoStop}%
\bibitem [{\citenamefont {Luca}\ and\ \citenamefont
  {Scardicchio}(2013)}]{Luca13}%
  \BibitemOpen
  \bibfield  {author} {\bibinfo {author} {\bibfnamefont {A.~De}\ \bibnamefont
  {Luca}}\ and\ \bibinfo {author} {\bibfnamefont {A.}~\bibnamefont
  {Scardicchio}},\ }\bibfield  {title} {\enquote {\bibinfo {title} {Ergodicity
  breaking in a model showing many-body localization},}\ }\href
  {http://stacks.iop.org/0295-5075/101/i=3/a=37003} {\bibfield  {journal}
  {\bibinfo  {journal} {EPL (Europhysics Letters)}\ }\textbf {\bibinfo {volume}
  {101}},\ \bibinfo {pages} {37003} (\bibinfo {year} {2013})}\BibitemShut
  {NoStop}%
\bibitem [{\citenamefont {Majumdar}\ \emph {et~al.}(2008)\citenamefont
  {Majumdar}, \citenamefont {Bohigas},\ and\ \citenamefont
  {Lakshminarayan}}]{Majumdar2008}%
  \BibitemOpen
  \bibfield  {author} {\bibinfo {author} {\bibfnamefont {Satya~N.}\
  \bibnamefont {Majumdar}}, \bibinfo {author} {\bibfnamefont {Oriol}\
  \bibnamefont {Bohigas}}, \ and\ \bibinfo {author} {\bibfnamefont {Arul}\
  \bibnamefont {Lakshminarayan}},\ }\bibfield  {title} {\enquote {\bibinfo
  {title} {Exact minimum eigenvalue distribution of an entangled random pure
  state},}\ }\href {\doibase 10.1007/s10955-008-9491-5} {\bibfield  {journal}
  {\bibinfo  {journal} {Journal of Statistical Physics}\ }\textbf {\bibinfo
  {volume} {131}},\ \bibinfo {pages} {33--49} (\bibinfo {year}
  {2008})}\BibitemShut {NoStop}%
\bibitem [{\citenamefont {Majumdar}(2011)}]{MajumdarBook}%
  \BibitemOpen
  \bibfield  {author} {\bibinfo {author} {\bibfnamefont {Satya~N.}\
  \bibnamefont {Majumdar}},\ }\href@noop {} {\emph {\bibinfo {title} {Extreme
  eigenvalues of Wishart matrices: application to entangled bipartite system,
  \emph{Oxford Handbook of Random Matrix Theory}}}}\ (\bibinfo  {publisher}
  {Oxford University Press},\ \bibinfo {year} {2011})\BibitemShut {NoStop}%
\bibitem [{\citenamefont {Kubotani}\ \emph {et~al.}(2008)\citenamefont
  {Kubotani}, \citenamefont {Adachi},\ and\ \citenamefont {Toda}}]{KAT2008}%
  \BibitemOpen
  \bibfield  {author} {\bibinfo {author} {\bibfnamefont {Hiroto}\ \bibnamefont
  {Kubotani}}, \bibinfo {author} {\bibfnamefont {Satoshi}\ \bibnamefont
  {Adachi}}, \ and\ \bibinfo {author} {\bibfnamefont {Mikito}\ \bibnamefont
  {Toda}},\ }\bibfield  {title} {\enquote {\bibinfo {title} {Exact formula of
  the distribution of schmidt eigenvalues for dynamical formation of
  entanglement in quantum chaos},}\ }\href {\doibase
  10.1103/PhysRevLett.100.240501} {\bibfield  {journal} {\bibinfo  {journal}
  {Phys. Rev. Lett.}\ }\textbf {\bibinfo {volume} {100}},\ \bibinfo {pages}
  {240501} (\bibinfo {year} {2008})}\BibitemShut {NoStop}%
\bibitem [{\citenamefont {Vivo}(2011)}]{Vivo2011}%
  \BibitemOpen
  \bibfield  {author} {\bibinfo {author} {\bibfnamefont {Pierpaolo}\
  \bibnamefont {Vivo}},\ }\bibfield  {title} {\enquote {\bibinfo {title}
  {Largest schmidt eigenvalue of random pure states and conductance
  distribution in chaotic cavities},}\ }\href {\doibase
  10.1088/1742-5468/2011/01/p01022} {\bibfield  {journal} {\bibinfo  {journal}
  {Journal of Statistical Mechanics: Theory and Experiment}\ }\textbf {\bibinfo
  {volume} {2011}},\ \bibinfo {pages} {P01022} (\bibinfo {year}
  {2011})}\BibitemShut {NoStop}%
\bibitem [{\citenamefont {Kumar}\ \emph {et~al.}(2017)\citenamefont {Kumar},
  \citenamefont {Sambasivam},\ and\ \citenamefont {Anand}}]{Kumar2017}%
  \BibitemOpen
  \bibfield  {author} {\bibinfo {author} {\bibfnamefont {Santosh}\ \bibnamefont
  {Kumar}}, \bibinfo {author} {\bibfnamefont {Bharath}\ \bibnamefont
  {Sambasivam}}, \ and\ \bibinfo {author} {\bibfnamefont {Shashank}\
  \bibnamefont {Anand}},\ }\bibfield  {title} {\enquote {\bibinfo {title}
  {Smallest eigenvalue density for regular or fixed-trace complex
  wishart{\textendash}laguerre ensemble and entanglement in coupled kicked
  tops},}\ }\href {\doibase 10.1088/1751-8121/aa7d0e} {\bibfield  {journal}
  {\bibinfo  {journal} {Journal of Physics A: Mathematical and Theoretical}\
  }\textbf {\bibinfo {volume} {50}},\ \bibinfo {pages} {345201} (\bibinfo
  {year} {2017})}\BibitemShut {NoStop}%
\bibitem [{\citenamefont {Forrester}\ and\ \citenamefont
  {Kumar}(2019)}]{Forrester2019}%
  \BibitemOpen
  \bibfield  {author} {\bibinfo {author} {\bibfnamefont {Peter~J}\ \bibnamefont
  {Forrester}}\ and\ \bibinfo {author} {\bibfnamefont {Santosh}\ \bibnamefont
  {Kumar}},\ }\bibfield  {title} {\enquote {\bibinfo {title} {Recursion scheme
  for the largest {\textdollar}{\textbackslash}beta{\textdollar}
  -wishart{\textendash}laguerre eigenvalue and landauer conductance in quantum
  transport},}\ }\href {\doibase 10.1088/1751-8121/ab433c} {\bibfield
  {journal} {\bibinfo  {journal} {Journal of Physics A: Mathematical and
  Theoretical}\ }\textbf {\bibinfo {volume} {52}},\ \bibinfo {pages} {42LT02}
  (\bibinfo {year} {2019})}\BibitemShut {NoStop}%
\bibitem [{\citenamefont {Tomsovic}\ \emph {et~al.}(2018)\citenamefont
  {Tomsovic}, \citenamefont {Lakshminarayan}, \citenamefont {Srivastava},\ and\
  \citenamefont {B\"acker}}]{Tomsovic2018}%
  \BibitemOpen
  \bibfield  {author} {\bibinfo {author} {\bibfnamefont {Steven}\ \bibnamefont
  {Tomsovic}}, \bibinfo {author} {\bibfnamefont {Arul}\ \bibnamefont
  {Lakshminarayan}}, \bibinfo {author} {\bibfnamefont {Shashi C.~L.}\
  \bibnamefont {Srivastava}}, \ and\ \bibinfo {author} {\bibfnamefont {Arnd}\
  \bibnamefont {B\"acker}},\ }\bibfield  {title} {\enquote {\bibinfo {title}
  {Eigenstate entanglement between quantum chaotic subsystems: Universal
  transitions and power laws in the entanglement spectrum},}\ }\href {\doibase
  10.1103/PhysRevE.98.032209} {\bibfield  {journal} {\bibinfo  {journal} {Phys.
  Rev. E}\ }\textbf {\bibinfo {volume} {98}},\ \bibinfo {pages} {032209}
  (\bibinfo {year} {2018})}\BibitemShut {NoStop}%
\bibitem [{\citenamefont {Lakshminarayan}\ \emph {et~al.}(2008)\citenamefont
  {Lakshminarayan}, \citenamefont {Tomsovic}, \citenamefont {Bohigas},\ and\
  \citenamefont {Majumdar}}]{AL08}%
  \BibitemOpen
  \bibfield  {author} {\bibinfo {author} {\bibfnamefont {Arul}\ \bibnamefont
  {Lakshminarayan}}, \bibinfo {author} {\bibfnamefont {Steven}\ \bibnamefont
  {Tomsovic}}, \bibinfo {author} {\bibfnamefont {Oriol}\ \bibnamefont
  {Bohigas}}, \ and\ \bibinfo {author} {\bibfnamefont {Satya~N.}\ \bibnamefont
  {Majumdar}},\ }\bibfield  {title} {\enquote {\bibinfo {title} {Extreme
  statistics of complex random and quantum chaotic states},}\ }\href {\doibase
  10.1103/PhysRevLett.100.044103} {\bibfield  {journal} {\bibinfo  {journal}
  {Phys. Rev. Lett.}\ }\textbf {\bibinfo {volume} {100}},\ \bibinfo {pages}
  {044103} (\bibinfo {year} {2008})}\BibitemShut {NoStop}%
\bibitem [{\citenamefont {Johnstone}(2001)}]{Johnstone01}%
  \BibitemOpen
  \bibfield  {author} {\bibinfo {author} {\bibfnamefont {Iain~M.}\ \bibnamefont
  {Johnstone}},\ }\bibfield  {title} {\enquote {\bibinfo {title} {On the
  distribution of the largest eigenvalue in principal components analysis},}\
  }\href {http://www.jstor.org/stable/2674106} {\bibfield  {journal} {\bibinfo
  {journal} {The Annals of Statistics}\ }\textbf {\bibinfo {volume} {29}},\
  \bibinfo {pages} {295--327} (\bibinfo {year} {2001})}\BibitemShut {NoStop}%
\bibitem [{\citenamefont {Nielsen}\ and\ \citenamefont
  {Chuang}(2011)}]{Nielsen11}%
  \BibitemOpen
  \bibfield  {author} {\bibinfo {author} {\bibfnamefont {Michael~A.}\
  \bibnamefont {Nielsen}}\ and\ \bibinfo {author} {\bibfnamefont {Isaac~L.}\
  \bibnamefont {Chuang}},\ }\href@noop {} {\emph {\bibinfo {title} {Quantum
  Computation and Quantum Information: 10th Anniversary Edition}}},\ \bibinfo
  {edition} {10th}\ ed.\ (\bibinfo  {publisher} {Cambridge University Press},\
  \bibinfo {address} {New York, NY, USA},\ \bibinfo {year} {2011})\BibitemShut
  {NoStop}%
\bibitem [{\citenamefont {Zyczkowski}\ and\ \citenamefont
  {Sommers}(2001)}]{SommersZyczkowski2001}%
  \BibitemOpen
  \bibfield  {author} {\bibinfo {author} {\bibfnamefont {Karol}\ \bibnamefont
  {Zyczkowski}}\ and\ \bibinfo {author} {\bibfnamefont {Hans-Jürgen}\
  \bibnamefont {Sommers}},\ }\bibfield  {title} {\enquote {\bibinfo {title}
  {Induced measures in the space of mixed quantum states},}\ }\href {\doibase
  10.1088/0305-4470/34/35/335} {\bibfield  {journal} {\bibinfo  {journal}
  {Journal of Physics A: Mathematical and General}\ }\textbf {\bibinfo {volume}
  {34}},\ \bibinfo {pages} {7111--7125} (\bibinfo {year} {2001})}\BibitemShut
  {NoStop}%
\bibitem [{\citenamefont {Bengtsson}\ and\ \citenamefont
  {{\.Z}yczkowski}(2017)}]{BZBook}%
  \BibitemOpen
  \bibfield  {author} {\bibinfo {author} {\bibfnamefont {I.}~\bibnamefont
  {Bengtsson}}\ and\ \bibinfo {author} {\bibfnamefont {K.}~\bibnamefont
  {{\.Z}yczkowski}},\ }\href {https://books.google.co.in/books?id=OD0yDwAAQBAJ}
  {\emph {\bibinfo {title} {Geometry of Quantum States: An Introduction to
  Quantum Entanglement}}}\ (\bibinfo  {publisher} {Cambridge University
  Press},\ \bibinfo {year} {2017})\BibitemShut {NoStop}%
\bibitem [{\citenamefont {Ginibre}(1965)}]{Ginibre65}%
  \BibitemOpen
  \bibfield  {author} {\bibinfo {author} {\bibfnamefont {Jean}\ \bibnamefont
  {Ginibre}},\ }\bibfield  {title} {\enquote {\bibinfo {title} {Statistical
  ensembles of complex, quaternion, and real matrices},}\ }\href {\doibase
  10.1063/1.1704292} {\bibfield  {journal} {\bibinfo  {journal} {J.
  Mathematical Phys.}\ }\textbf {\bibinfo {volume} {6}},\ \bibinfo {pages}
  {440--449} (\bibinfo {year} {1965})}\BibitemShut {NoStop}%
\bibitem [{\citenamefont {Tao}\ and\ \citenamefont {Vu}(2012)}]{TaoVu2012}%
  \BibitemOpen
  \bibfield  {author} {\bibinfo {author} {\bibfnamefont {Terence}\ \bibnamefont
  {Tao}}\ and\ \bibinfo {author} {\bibfnamefont {Van}\ \bibnamefont {Vu}},\
  }\bibfield  {title} {\enquote {\bibinfo {title} {Random covariance matrices:
  Universality of local statistics of eigenvalues},}\ }\href {\doibase
  10.1214/11-AOP648} {\bibfield  {journal} {\bibinfo  {journal} {Ann. Probab.}\
  }\textbf {\bibinfo {volume} {40}},\ \bibinfo {pages} {1285--1315} (\bibinfo
  {year} {2012})}\BibitemShut {NoStop}%
\bibitem [{\citenamefont {Haake}(2001)}]{Haake2001}%
  \BibitemOpen
  \bibfield  {author} {\bibinfo {author} {\bibfnamefont {F.}~\bibnamefont
  {Haake}},\ }\href {https://books.google.co.kr/books?id=Orv0BXoorFEC} {\emph
  {\bibinfo {title} {Quantum Signatures of Chaos}}},\ Physics and astronomy
  online library\ (\bibinfo  {publisher} {Springer},\ \bibinfo {year}
  {2001})\BibitemShut {NoStop}%
\bibitem [{\citenamefont {Nechita}(2007)}]{Nechita07}%
  \BibitemOpen
  \bibfield  {author} {\bibinfo {author} {\bibfnamefont {Ion}\ \bibnamefont
  {Nechita}},\ }\bibfield  {title} {\enquote {\bibinfo {title} {Asymptotics of
  random density matrices},}\ }\href {\doibase 10.1007/s00023-007-0345-5}
  {\bibfield  {journal} {\bibinfo  {journal} {Annales Henri Poincar{\'e}}\
  }\textbf {\bibinfo {volume} {8}},\ \bibinfo {pages} {1521--1538} (\bibinfo
  {year} {2007})}\BibitemShut {NoStop}%
\bibitem [{\citenamefont {Luitz}\ \emph {et~al.}(2015)\citenamefont {Luitz},
  \citenamefont {Laflorencie},\ and\ \citenamefont {Alet}}]{Luitz15}%
  \BibitemOpen
  \bibfield  {author} {\bibinfo {author} {\bibfnamefont {David~J.}\
  \bibnamefont {Luitz}}, \bibinfo {author} {\bibfnamefont {Nicolas}\
  \bibnamefont {Laflorencie}}, \ and\ \bibinfo {author} {\bibfnamefont
  {Fabien}\ \bibnamefont {Alet}},\ }\bibfield  {title} {\enquote {\bibinfo
  {title} {Many-body localization edge in the random-field heisenberg chain},}\
  }\href {\doibase 10.1103/PhysRevB.91.081103} {\bibfield  {journal} {\bibinfo
  {journal} {Phys. Rev. B}\ }\textbf {\bibinfo {volume} {91}},\ \bibinfo
  {pages} {081103} (\bibinfo {year} {2015})}\BibitemShut {NoStop}%
\bibitem [{\citenamefont {Regnault}\ and\ \citenamefont
  {Nandkishore}(2016)}]{Nandkishore16}%
  \BibitemOpen
  \bibfield  {author} {\bibinfo {author} {\bibfnamefont {Nicolas}\ \bibnamefont
  {Regnault}}\ and\ \bibinfo {author} {\bibfnamefont {Rahul}\ \bibnamefont
  {Nandkishore}},\ }\bibfield  {title} {\enquote {\bibinfo {title} {Floquet
  thermalization: Symmetries and random matrix ensembles},}\ }\href {\doibase
  10.1103/PhysRevB.93.104203} {\bibfield  {journal} {\bibinfo  {journal} {Phys.
  Rev. B}\ }\textbf {\bibinfo {volume} {93}},\ \bibinfo {pages} {104203}
  (\bibinfo {year} {2016})}\BibitemShut {NoStop}%
\bibitem [{\citenamefont {Brody}\ \emph {et~al.}(1981)\citenamefont {Brody},
  \citenamefont {Flores}, \citenamefont {French}, \citenamefont {Mello},
  \citenamefont {Pandey},\ and\ \citenamefont {Wong}}]{Pandey81}%
  \BibitemOpen
  \bibfield  {author} {\bibinfo {author} {\bibfnamefont {T.~A.}\ \bibnamefont
  {Brody}}, \bibinfo {author} {\bibfnamefont {J.}~\bibnamefont {Flores}},
  \bibinfo {author} {\bibfnamefont {J.~B.}\ \bibnamefont {French}}, \bibinfo
  {author} {\bibfnamefont {P.~A.}\ \bibnamefont {Mello}}, \bibinfo {author}
  {\bibfnamefont {A.}~\bibnamefont {Pandey}}, \ and\ \bibinfo {author}
  {\bibfnamefont {S.~S.~M.}\ \bibnamefont {Wong}},\ }\bibfield  {title}
  {\enquote {\bibinfo {title} {Random-matrix physics: spectrum and strength
  fluctuations},}\ }\href {\doibase 10.1103/RevModPhys.53.385} {\bibfield
  {journal} {\bibinfo  {journal} {Rev. Mod. Phys.}\ }\textbf {\bibinfo {volume}
  {53}},\ \bibinfo {pages} {385--479} (\bibinfo {year} {1981})}\BibitemShut
  {NoStop}%
\bibitem [{\citenamefont {Johnstone}\ \emph {et~al.}(2014)\citenamefont
  {Johnstone}, \citenamefont {Ma}, \citenamefont {Perry},\ and\ \citenamefont
  {Shahram}}]{Rmt14}%
  \BibitemOpen
  \bibfield  {author} {\bibinfo {author} {\bibfnamefont {Iain~M.}\ \bibnamefont
  {Johnstone}}, \bibinfo {author} {\bibfnamefont {Zongming}\ \bibnamefont
  {Ma}}, \bibinfo {author} {\bibfnamefont {Patrick~O.}\ \bibnamefont {Perry}},
  \ and\ \bibinfo {author} {\bibfnamefont {Morteza}\ \bibnamefont {Shahram}},\
  }\href@noop {} {\emph {\bibinfo {title} {RMTstat: Distributions, Statistics
  and Tests derived from Random Matrix Theory}}} (\bibinfo {year} {2014}),\
  \bibinfo {note} {r package version 0.3}\BibitemShut {NoStop}%
\bibitem [{\citenamefont {Luitz}(2016)}]{Luitz16}%
  \BibitemOpen
  \bibfield  {author} {\bibinfo {author} {\bibfnamefont {David~J.}\
  \bibnamefont {Luitz}},\ }\bibfield  {title} {\enquote {\bibinfo {title} {Long
  tail distributions near the many-body localization transition},}\ }\href
  {\doibase 10.1103/PhysRevB.93.134201} {\bibfield  {journal} {\bibinfo
  {journal} {Phys. Rev. B}\ }\textbf {\bibinfo {volume} {93}},\ \bibinfo
  {pages} {134201} (\bibinfo {year} {2016})}\BibitemShut {NoStop}%
\bibitem [{\citenamefont {Lim}\ and\ \citenamefont {Sheng}(2016)}]{limsheng16}%
  \BibitemOpen
  \bibfield  {author} {\bibinfo {author} {\bibfnamefont {S.~P.}\ \bibnamefont
  {Lim}}\ and\ \bibinfo {author} {\bibfnamefont {D.~N.}\ \bibnamefont
  {Sheng}},\ }\bibfield  {title} {\enquote {\bibinfo {title} {Many-body
  localization and transition by density matrix renormalization group and exact
  diagonalization studies},}\ }\href {\doibase 10.1103/PhysRevB.94.045111}
  {\bibfield  {journal} {\bibinfo  {journal} {Phys. Rev. B}\ }\textbf {\bibinfo
  {volume} {94}},\ \bibinfo {pages} {045111} (\bibinfo {year}
  {2016})}\BibitemShut {NoStop}%
\bibitem [{\citenamefont {Mishra}\ and\ \citenamefont
  {Lakshminarayan}(2014)}]{Mishra2014}%
  \BibitemOpen
  \bibfield  {author} {\bibinfo {author} {\bibfnamefont {Sunil~K.}\
  \bibnamefont {Mishra}}\ and\ \bibinfo {author} {\bibfnamefont {Arul}\
  \bibnamefont {Lakshminarayan}},\ }\bibfield  {title} {\enquote {\bibinfo
  {title} {Resonance and generation of random states in a quenched ising
  model},}\ }\href {\doibase 10.1209/0295-5075/105/10002} {\bibfield  {journal}
  {\bibinfo  {journal} {{EPL} (Europhysics Letters)}\ }\textbf {\bibinfo
  {volume} {105}},\ \bibinfo {pages} {10002} (\bibinfo {year}
  {2014})}\BibitemShut {NoStop}%
\bibitem [{\citenamefont {Buijsman}\ \emph {et~al.}(2019)\citenamefont
  {Buijsman}, \citenamefont {Gritsev},\ and\ \citenamefont
  {Cheianov}}]{Gumbel19}%
  \BibitemOpen
  \bibfield  {author} {\bibinfo {author} {\bibfnamefont {Wouter}\ \bibnamefont
  {Buijsman}}, \bibinfo {author} {\bibfnamefont {Vladimir}\ \bibnamefont
  {Gritsev}}, \ and\ \bibinfo {author} {\bibfnamefont {Vadim}\ \bibnamefont
  {Cheianov}},\ }\bibfield  {title} {\enquote {\bibinfo {title} {Gumbel
  statistics for entanglement spectra of many-body localized eigenstates},}\
  }\href {\doibase 10.1103/PhysRevB.100.205110} {\bibfield  {journal} {\bibinfo
   {journal} {Phys. Rev. B}\ }\textbf {\bibinfo {volume} {100}},\ \bibinfo
  {pages} {205110} (\bibinfo {year} {2019})}\BibitemShut {NoStop}%
\end{thebibliography}%

\end{document}